\documentclass[preprints,review,accept,moreauthors]{mdpi} 
\usepackage{mathrsfs,amsmath} 
 \usepackage{setspace}
 \usepackage{hyperref}
\firstpage{1} 
\pubvolume{xx}
\issuenum{1}
\articlenumber{5}
\pubyear{2019}
\copyrightyear{2019}
\history{\href{https://www.mdpi.com/2072-666X/11/2/230/htm}{Published in Micromachines {\bf 11},(2) 230 (2020)}}

\pdfoutput=1

\Title{Optical frequency combs in quadratically nonlinear resonators}

\Author{
Iolanda~Ricciardi$^{1,2}$,
Simona Mosca$^{1}$,
Maria Parisi$^{1}$,
François Leo$^{3}$, 
Tobias Hansson$^{4}$,
\\
Miro Erkintalo$^{5,6}$,
Pasquale Maddaloni$^{1,2}$, 
Paolo De~Natale$^{7}$
Stefan Wabnitz$^{1,8,9}$,
\\
and Maurizio De~Rosa$^{1,2}$* 
}

\AuthorNames{Firstname Lastname, Firstname Lastname and Firstname Lastname}

\address{%
$^{1}$ \quad CNR-INO, Istituto Nazionale di Ottica, Via Campi Flegrei 34, I-80078 Pozzuoli (NA), Italy;\\
$^{2}$ \quad INFN, Istituto Nazionale di Fisica Nucleare, Sez. di Napoli, Complesso Universitario di M.S. Angelo, Via Cintia, Napoli 80126, Italy;\\
$^{3}$ \quad OPERA-photonics, Universit\'e libre de Bruxelles, 50 Avenue F. D. Roosevelt, CP 194/5, B-1050 Bruxelles, Belgium;\\
$^{4}$ \quad Department of Physics, Chemistry and Biology, Link\"oping University, SE-581 83 Link\"oping, Sweden;\\
$^{5}$ \quad The Dodd-Walls Centre for Photonic and Quantum Technologies, Auckland 1142, New Zealand\\
$^{6}$ \quad Physics Department, The University of Auckland, Auckland 1142, New Zealand;\\
$^{7}$ \quad CNR-INO, Istituto Nazionale di Ottica, Largo E. Fermi 6, I-50125 Firenze, Italy; \\
$^{8}$ \quad Dipartimento di Ingegneria dell'Informazione, Elettronica e Telecomunicazioni, Sapienza Universit\`a di Roma- Via Eudossiana 18,
I-00184 Roma, Italy;\\
$^{9}$ \quad Department of Physics, Novosibirsk State University, 1 Pirogova Street, Novosibirsk 630090, Russia;
}

\corres{Correspondence: maurizio.derosa@ino.cnr.it}

\abstract{
Optical frequency combs are one of the most remarkable inventions of the last decades. 
Originally conceived as the spectral counterpart of the train of short pulses emitted by mode-locked lasers, frequency combs have also been subsequently generated in continuously pumped microresonators, through third-order parametric processes. 
Quite recently, direct generation of optical frequency combs has been demonstrated in continuous-wave laser-pumped optical resonators with a second-order nonlinear medium inside. 
Here, we presents a concise introduction to such quadratic combs and the physical mechanism that underlies their formation. We mainly review our recent experimental and theoretical work on formation and dynamics of such quadratic frequency combs.
We experimentally demonstrated comb generation in two configurations: a cavity for second harmonic generation, where combs are generated both around the pump frequency and  its second harmonic, and a degenerate optical parametric oscillator, where combs are generated around the pump frequency and  its sub-harmonic. 
The experiments have been supported by a thorough theoretical analysis, aimed at modelling the dynamics of quadratic combs, both in frequency and time domains, providing useful insights into the physics of this new class of optical frequency comb synthesizers. 
Quadratic combs establish a new class of efficient frequency comb synthesizers, with unique features, which could enable straightforward access to new spectral regions and stimulate novel applications.   
}


\begin{document}

\section{Introduction}
Twenty years ago, optical frequency combs (OFCs) were established as powerful tools for accurate measurements of optical frequencies and timekeeping~\cite{Jones:2000tn,Holzwarth:2000aa},  a result of a long-standing effort, which was recognized with the Nobel Prize in Physics in 2005~\cite{Hall:2006by,Hansch:2006el}. 
The discrete ensemble of equally spaced laser frequencies that distinguish OFCs from other light sources is the spectral counterpart of the regular train of short pulses emitted by mode-locked lasers, which were initially used for comb generation. 
OFCs have become a critical component in many scientific and technological  applications~\cite{Newbury:2011dh}, from highly accurate optical frequency measurements for fundamental tests of physics~\cite{Predehl:2012kr,Clivati:2016ij,Insero:2017gg,DiSarno:2019ep} to exoplanet exploration~\cite{Steinmetz:2008he,McCracken:2017dy,Obrzud:2018cb}\, from air pollution detection~\cite{Adler:2010da,Keilmann:2004gd,Picque:2019jz,Schliesser:2012dn,Rieker:2014fj,Yu:2018ku} to telecommunication systems~\cite{Pfeifle:2014cm,Kemal:2016ho,MarinPalomo:2017bv}, while a growing interest have aroused in the quantum properties of OFCs~\cite{Roslund:2014cb,Dutt:2015fx,Reimer:2016jk,Imany:2018jk,Kues:2019dh}.

Thereafter, comb emission was also demonstrated in continuous-wave (cw) laser-pumped resonators through cascaded third-order $\chi^{(3)}$ parametric processes~\cite{DelHaye:2007gi}. 
In such Kerr resonators, a first pair of sidebands is generated around the pump frequency by cavity modulation instability or degenerate four-wave mixing (FWM); subsequently, cascaded four-wave mixing processes lead to the  formation, around the pump frequency, of a uniform frequency comb, where self- and cross-phase modulation act to compensate for the unequal cavity mode spacing induced by the group velocity dispersion (GVD)~\cite{Kippenberg:2011fc,Pasquazi:2018de}. 
Because of the relatively low strength of third-order nonlinearity, generation of Kerr combs requires small interaction volumes and high-$Q$ resonators. For these reasons, small size resonators are particularly suited to reach broadband comb generation with quite moderate pump power~\cite{Gaeta:2019hc}. Moreover, when the mode size is comparable with the light wavelength, a careful design of the resonator geometry can effectively modify the GVD of the resonator, leading to a broader spectral emission.

While $\chi^{(2)}$ three-wave mixing processes have been widely used for spectral conversion of femtosecond laser combs since their inception~\cite{Maddaloni:2006ka,Sun:2007wu,Wong:2008fz,Gambetta:2008fa,Adler:2009ka,Galli:2013cg}, only in the last years it was demonstrated that quadratic $\chi^{(2)}$ processes can lead to direct generation of optical frequency combs in cw-pumped quadratic nonlinear resonator.    
Actually, in 1999 Diddams at al.  generated an OFC in a second-order nonlinear system, by actively inducing intracavity phase modulation inside a cw-pumped nearly degenerate optical parametric oscillator (OPO)~\cite{Diddams:1999ch}, following a long development of phase modulation in lithium niobate for comb generation~\cite{Kourogi:1993gy}.
 According to this scheme, besides the nonlinear crystal for parametric amplification, a phase modulator was placed inside the OPO cavity and driven at a modulation frequency equal to the cavity free spectral range. The modulator thus generated a family of phase-coupled sidebands, around the nearly degenerate signal and idler waves, which coincided with the resonator mode frequencies.  
Unlike other works presented in the following, where combs arise through purely $\chi^{(2)}$ optical processes, in that work combs were initially seeded by the sidebands generated in the intracavity modulator.
Optical parametric amplification further increased the number of resonant sidebands, eventually leading to a 18-nm wide comb of equally spaced, mode-locked lines around the degenerate OPO frequency, only limited by the dispersive shift of the cavity modes, where mode-locking is imposed by phase modulation.

More recently, an optical frequency comb was produced by adding a second nonlinear crystal in a nondegenerate OPO~\cite{Ulvila:2013jv}. 
The authors observed comb formation around the signal wavelength when the second crystal was phase mismatched for second harmonic generation (SHG) of the signal wave. Subsequent investigations of the same system reported experimental evidence of a comb around the second harmonic of the signal wave, whereas the comb around the signal was simultaneously transferred to the idler spectral range by parametric amplification~\cite{Ulvila:2014bx}. 
 In this case, the phase mismatched crystal behaves like a Kerr medium, producing a phase shift of the fundamental wave, which is proportional to the field intensity~\cite{Ostrovskii:1967,Desalvo:1992cs,Stegeman:1999fe}. 
This phase shift can be explained as the consequence of cascaded quadratic processes which occur in the crystal when SHG is not phase matched.
Indeed, when the fundamental pump wave, at frequency $\omega/2\pi$ enters a nonlinear crystal, a second harmonic field is generated, $\omega + \omega \rightarrow  2\omega$. If the process is not phase matched, the second harmonic (SH) field travels at a different phase velocity and, after half a coherence length, down-converts back to the fundamental frequency, $2\omega-\omega \rightarrow\omega$, with a different phase from that of the unconverted pump field.

As we will see later, a different cascaded three-wave-mixing process is decisive for the onset of frequency combs in phase-matched intracavity SHG---namely, internally pumped optical parametric oscillation~\cite{Schiller:1993tz,Schiller:1996gx}. 
In fact, degenerate optical parametric oscillation and SHG are mutually inverse processes, which satisfy the same phase matching condition, $\Delta k=2k_1-k_2=0$, between wave vectors $k_1=k(\omega)$ of the fundamental field and $k_2=k(2\omega)$ of the second harmonic field, respectively. Therefore, a properly phase-matched crystal placed inside an optical resonator can work either for SHG or parametric oscillation, depending on whether it is pumped at the fundamental or second harmonic frequency, respectively. 
However, the harmonic field generated in the first case can act as a pump for a nondegenerate cascaded OPO, and a pair of parametric fields start to oscillate with frequencies symmetrically placed around the fundamental pump.
Although internally pumped OPO was observed and investigated for a long time, before the importance of OFCs was established~\cite{Schiller:1993tz,Schiller:1996gx,Schneider:1997ks,White:1997ta}, the observation of frequency combs in quadratic nonlinear media was postponed to recent years.

Here, we present a  concise introduction to the physical mechanism that underlies quadratic comb formation, as well as an extended theoretical framework that has been developed so far.  
We particularly focus on our recent activity in this field, discussing our experimental and theoretical work on direct generation of quadratic combs. As a whole, it represents a systematic and coherent, although not exhaustive, approach to this new field.
After the work of Ref.~\cite{Ulvila:2013jv} , Ricciardi et al. experimentally demonstrated direct frequency comb generation in an optical resonator with a single nonlinear crystal inside, originally conceived for cavity-enhanced SHG. OFCs were observed in the case of both phase-matched and phase-mismatched SHG. Moreover, the authors presented a simple theoretical model, which explained comb generation as the result of cascaded $\chi^{(2)}\hspace{-4pt}:\hspace{-2pt}\chi^{(2)}$ processes~\cite{Ricciardi:2015bw,Mosca:2015wh}. 
A more general theoretical description of comb generation in SHG cavity was successively developed by Leo et al., who modeled the dynamics of the cavity field in the time domain~\cite{Leo:2016kj,Leo:2016df,Hansson:2017cs}, and described comb formation in the framework of a modulation instability (MI), i.e., the growth of sidebands around a carrier frequency by amplification of small modulations on the carrier wave~\cite{Zakharov:2009du}. A similar theoretical description was adopted to describe the dynamics of quadratic combs observed in a degenerate OPO~\cite{Mosca:2018jk}. Finally, the most general approach, based on a single-envelope equation, has been also developed in order to study multi-octave, quadratic comb formation~\cite{Hansson:2016kz}.

Quadratically nonlinear resonators thus emerge as the basis of an entirely new class of highly efficient synthesizers of OFCs, with unique features, such as the simultaneous generation of frequency combs in spectral regions far from the pump frequency, and the role of phase matching in mitigating the effect of dispersion. Compared to Kerr combs, quadratic combs exploit the intrinsically higher efficiency of second-order nonlinearity, reducing the requirement in terms of pump power. 
Quadratic combs are still at an early stage but they are attracting the interest of an increasing number of research groups. 
More recent works are briefly reviewed in Section 6,  where we conclude by giving an overview of promising developments of quadratic combs in terms of material platforms for chip-scale devices, steady low-noise dynamical regimes, and their potential interest for quantum optics.

\section{Intracavity second harmonic generation}

\begin{figure}[t!]
\centering
\includegraphics[width=13 cm]{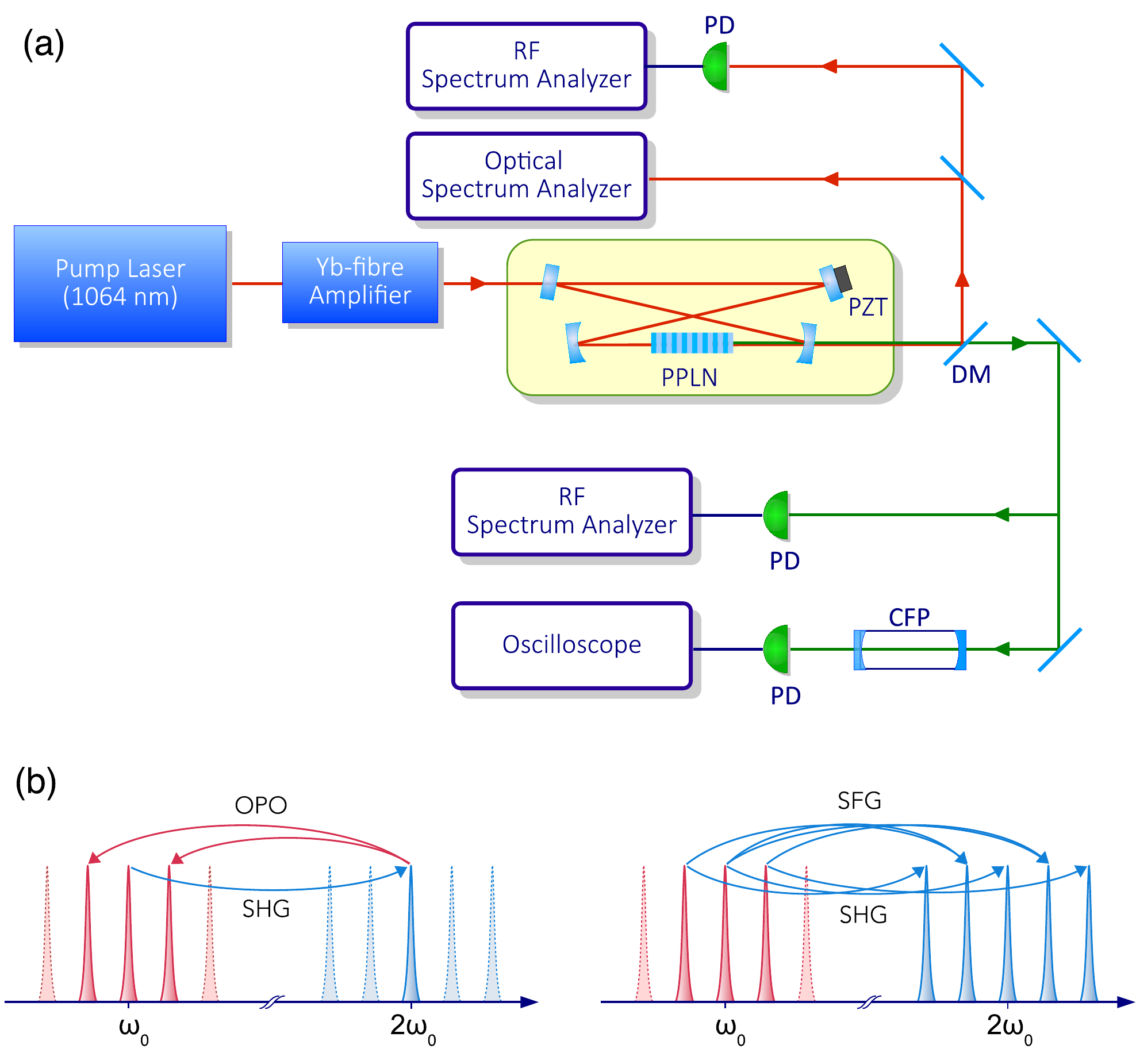}
\caption{Singly resonant cavity SHG. (a) Experimental setup: periodically poled lithium niobate crystal (PPLN), piezoelectric actuator (PZT), photodiode (PD), dichroic mirror (DM). The output beams are detected and processed by radio-frequency (RF) analyzers, while optical spectral analysis is performed by an optical spectrum analyzer in the infrared range and a confocal Fabry-P\`erot interferometer (CFP) in the visible range. (b) Schematic representation of the first steps leading to the formation of a dual optical frequency comb in cavity-enhanced second-harmonic generation: (left) second-harmonic generation with cascaded nondegenerate OPO gives rise to two subharmonic sidebands, which in turn (right) lead to successive, multiple second-harmonic and sum-frequency generations. Adapted with permission from~\cite{Ricciardi:2015bw}. Copyrighted by the American Physical Society.}
\label{fig:SR-SHG}
\end{figure}

The first system that we investigated for the generation of quadratic OFCs was a cw-pumped, cavity enhanced SHG system. The system, shown in Fig.~\ref{fig:SR-SHG}(a), was based on a 15-mm-long periodically poled LiNbO$_3$ crystal, placed inside a traveling-wave optical resonator (free spectral range FSR=493~MHz, quality factor $Q=10^8$), resonating at the fundamental laser frequency $\omega_0$. Mirror reflectivities were chosen in order to facilitate the onset of an internally pumped OPO~\cite{Ricciardi:2015bw}. The crystal was pumped by a narrow-line, 1064-nm-wavelength Nd:YAG laser, amplified by a Yb-doped fiber amplifier. Frequency locking of a cavity resonance to the laser was achieved by the Pound-Drever-Hall technique~\cite{Drever:1983gx}.

\begin{figure}[t!]
\centering
\includegraphics[width=13 cm]{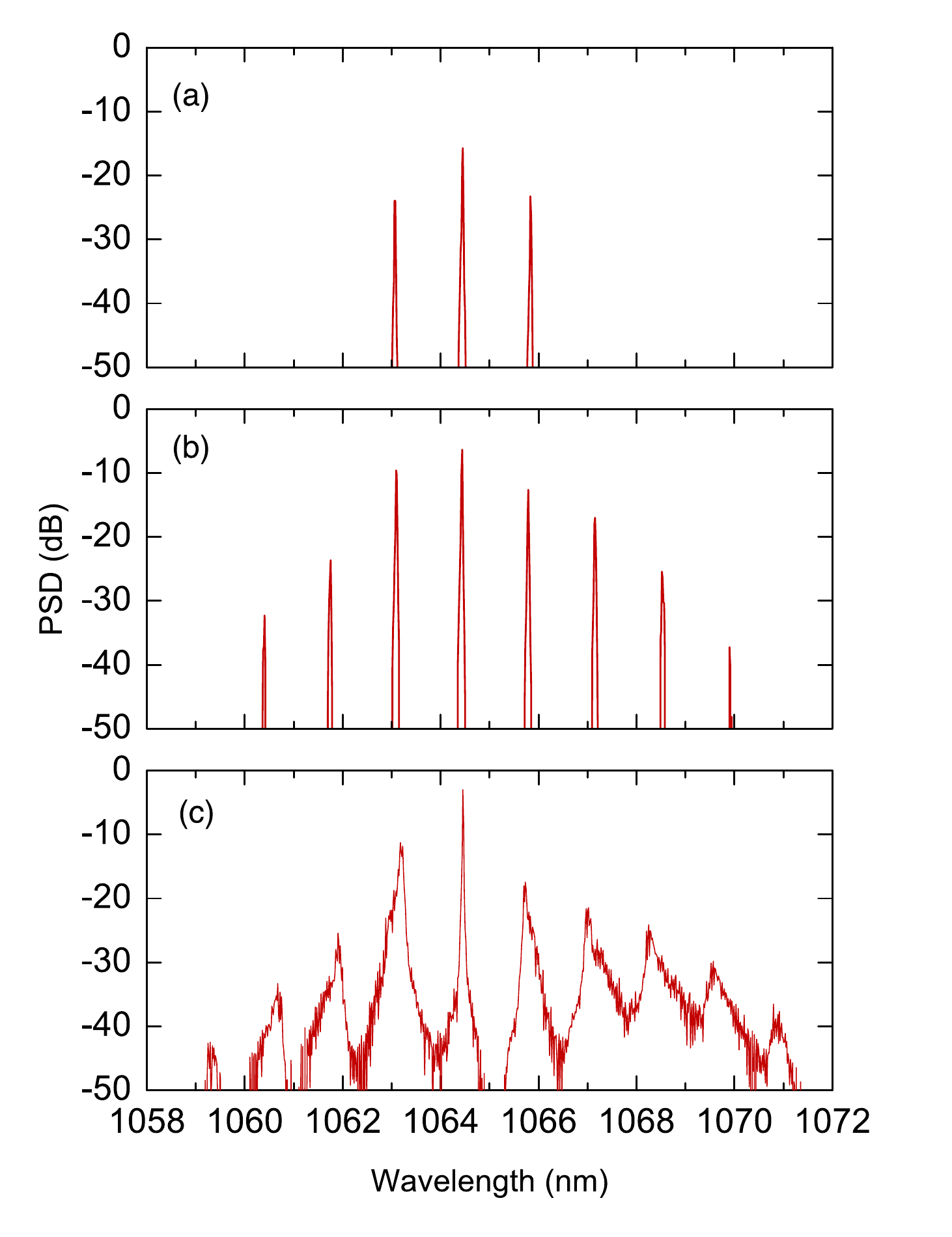}
\caption{Optical spectral power around the fundamental mode for (a) 170 mW, (b) 2 W, and (c) 9 W of input powers. Adapted with permission from~\cite{Ricciardi:2015bw}. Copyrighted by the American Physical Society.}
\label{fig:SR-PM}
\end{figure}

The phase-matching condition for SHG was achieved by properly adjusting the crystal temperature. Under this condition, we observed a first regime of pure harmonic generation, where the harmonic power increased with the input pump power. As shown in Fig.~\ref{fig:SR-PM}(a), when the input power exceeded the threshold for internally pumped OPO, the second harmonic power ceased to grow, and two parametric waves started to oscillate at frequencies $\omega_0\pm\Delta\omega$, symmetrically placed around the fundamental frequency (FF). 
As the power was further increased, additional sidebands appeared, displaced by multiples of $\Delta\omega$, leading to a multiple-FSR-spaced frequency comb, as sketched in Fig.~\ref{fig:SR-PM}(b). 
Finally, when the input power exceeded 5~W, secondary combs appeared around each of the primary comb lines, shown in Fig.~\ref{fig:SR-PM}(c). 
These secondary combs were spaced by 1 cavity FSR, as confirmed by the intermodal beat notes detected by fast photodetectors, both in the IR and in the visible spectral regions, and processed by a radio frequency (RF) spectrum analyzer.

\begin{figure}[t]
\centering
\includegraphics*[viewport= 0 280 380 660, clip,width=14 cm]{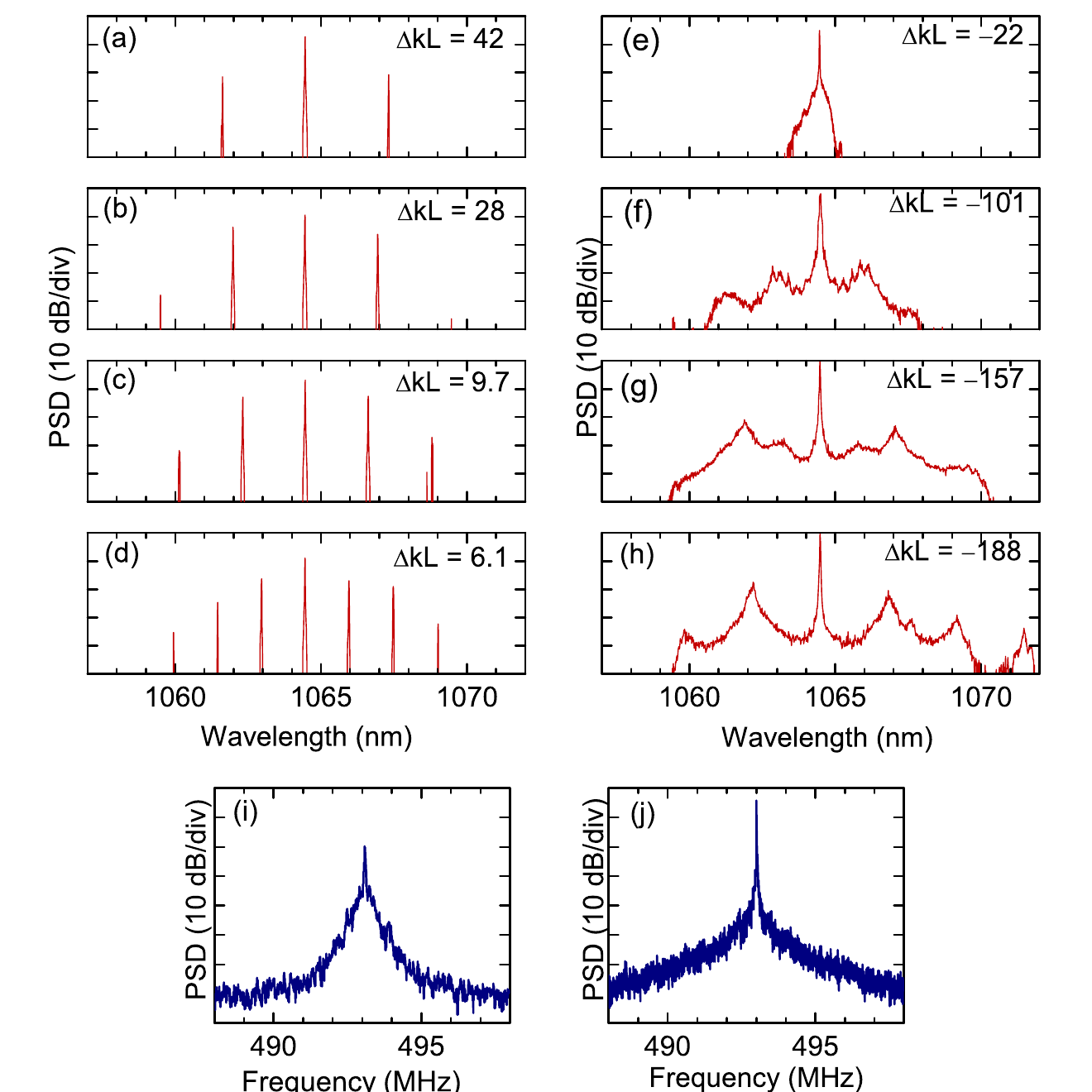}
\caption{Optical spectra for phase-mismatched singly resonant cavity SHG. (a)-(d) Positive phase mismatch; (e)-(h) negative phase mismatch. Intermodal beat notes corresponding to  (i) comb spectrum of Fig.~\ref{fig:SR-PM}(c), (j) comb spectrum in panel (g).} 
\label{fig:fuoriPM}
\end{figure}

Subsequently, wave vector mismatch $\Delta k$ was changed to finite values by varying the crystal temperature.
Figure \ref{fig:fuoriPM} shows infrared spectra observed for different values of the mismatch vector. For a positive mismatch, $\Delta k>0$, the spectra (a)-(d) show widely separated sidebands, similar to the spectra observed at $\Delta k=0$ (see Fig.~\ref{fig:SR-PM}(b)). The spacing between sidebands, as well as the pump power threshold for cascaded optical parametric oscillation, rapidly increases with the mismatch. 
For $\Delta k<0$, the spectra (e)-(h) consist of closely spaced (1 FSR) comb lines, and the spectral bandwidth increases with the magnitude of the mismatch. Larger negative phase mismatches are precluded by the limited accessible temperature range.
Figures \ref{fig:fuoriPM}(i) and (j) show the beat notes corresponding to the comb in Fig.~\ref{fig:SR-PM}(c) and the the comb in Fig.~\ref{fig:fuoriPM}(g), respectively.
The broad feature of the beat note (i) reveals a strong intermodal phase noise and, as a consequence, a low degree of coherence between the comb teeth. 
This feature is consistent with a scenario where comb modes are weakly coupled with each other, as they originate independently from each other.
On the contrary,  the beat note (j) is extremely narrow, being limited by the detection resolution bandwidth, and indicates a low intermodal phase noise and thus a strong phase coupling between all the comb teeth.

It is worth noting that the nonlinear resonator exhibits a noticeable thermal effect, mainly due to light absorption in the nonlinear crystal, which generates heat and leads to an increase of the cavity optical path, via thermal expansion and thermo-optic effect~\cite{Ricciardi:2010kd}. 
The photothermal effect introduces an additional nonlinear dynamical mechanism, with a temporal scale determined by the thermal diffusion time over the typical optical beam size~\cite{DeRosa:2002cg}. 
In our case, the photothermal effect was helpful in thermally locking a cavity resonance to the laser frequency~\cite{Carmon:2004us} when, specially at higher power, the PDH locking scheme was less effective.  
However, a better comprehension of the effect of thermal dynamics on comb formation requires further investigations.

As anticipated in the introduction, the onset of internally pumped OPO marks the beginning of a cascade of second-order nonlinear processes, which eventually produces a comb of equally spaced frequencies. As depicted in Fig.~\ref{fig:SR-SHG}(b), once generated, each parametric mode can generate new field modes through second harmonic, $(\omega + \Delta\omega) +(\omega + \Delta\omega) \rightarrow 2\omega + 2\Delta\omega$, and sum frequency with the fundamental wave, $\omega + (\omega + \Delta\omega) \rightarrow 2\omega + \Delta\omega$, processes, respectively. 
All these processes have been considered for the derivation of a simple system of coupled mode equations for the three intracavity subharmonic electric field amplitudes, the fundamental $A_0$, at $\omega_0$, and the parametric intracavity fields $A_\mu$ and $A_{\bar{\mu}}$, at $\omega_\mu=\omega_0+\Delta\omega$ and $\omega_{\bar{\mu}}=\omega_0-\Delta\omega$, respectively, which read~\cite{Ricciardi:2015bw}

\begin{align}
\dot{A}_0 =& 
-(\gamma_0 + i \Delta_0) \, A_0 
- 2 g \, \eta_{00\mu\bar{\mu}} A_0^* A_{\mu} A_{\bar{\mu}} 
- g (\eta_{0000} |A_0|^2
+2\eta_{0\mu0\mu} |A_{\mu}|^2
+2\eta_{0\bar{\mu}0\bar{\mu}} |A_{\bar{\mu}}|^2)  \, A_0  
+ F_\text{in}  
\label{eq:d-FWM-a}
\\
\dot{A}_{\mu} =& -(\gamma_{\mu} + i \Delta_{\mu}) \, A_{\mu}  
- g \, \eta_{\mu\bar{\mu}00} \, A_0^2 A^*_{\bar{\mu}} 
- g (2 \eta_{\mu00\mu} |A_0|^2 
+\eta_{\mu\mu\mu\mu} |A_{\mu}|^2 + 2\eta_{\mu\bar{\mu}\mu\bar{\mu}} |A_{\bar{\mu}}|^2)  \, A_{\mu}
\label{eq:d-FWM-b}
\\
\dot{A}_{\bar{\mu}} =& 
-(\gamma_{\bar{\mu}} + i \Delta_{\bar{\mu}}) \, A_{\bar{\mu}} 
- g \, \eta_{\bar{\mu}\mu00} \, A_0^2 A^*_{\mu}
- g (2 \eta_{\bar{\mu}00\bar{\mu}} |A_0|^2 
+ 2 \eta_{\bar{\mu}\mu\bar{\mu}\mu} |A_{\mu}|^2 
+ \eta_{\bar{\mu}\bar{\mu}\bar{\mu}\bar{\mu}} |A_{\bar{\mu}}|^2)  \, A_{\bar{\mu}} \, .
\label{eq:d-FWM-c}
\end{align}
Here, $F_\text{in}=\sqrt{2\gamma_0 /t_\text{R}}A_\text{in}$ is the cavity coupled amplitude of the constant input driving field $A_\text{in}$, at frequency $\omega_0$; the $\gamma$'s are the cavity decay constants; the $\Delta$'s are the cavity detunings of the respective modes; $g = (\kappa L)^2 /2t_\text{R}$ is a gain factor depending on the crystal length $L$ (hereafter we consider the cavity length equal to the crystal length); $t_\text{R}$ is the cavity round-trip time; and $\kappa=\sqrt{8}\omega_0 \chi^{(2)}_\mathrm{eff} /\sqrt{c^3 n_1^2  n_2 \epsilon_0}$ is the second-order coupling strength. The latter is normalized so that the square modulus of the field amplitudes is measured in watts, with $\chi^{(2)}_\mathrm{eff}$ the effective second-order susceptibility, $c$ the speed of light, $n_{1,2}$ the refractive indices, and $\epsilon_0$ the vacuum permittivity.
The integer mode number $\mu$ denotes the $\mu$th cavity mode, starting from the central mode at $\omega_0$, and overline stands for negative (lower frequencies). 
The $\eta$’s are complex nonlinear coupling constants, depending on the wave-vector mismatches associated with a pair of cascaded second-order processes,

\begin{equation}
\eta_{\mu\sigma\rho\nu} = \frac{2}{L^2} \int_{0}^{L}\!\!\! \int_{0}^{z} 
\exp{[-i (\xi_{\mu\sigma}z -\xi_{\rho\nu} z')]} \; \text{d}z' \, \text{d}z \
\label{eq:eta}
\end{equation}
where $\xi_{jk} =  k_{\omega_{j}} + k_{\omega_{k}} - k_{\omega_{j}+\omega_{k}}$.

A linear stability analysis of Eqs.~(\ref{eq:d-FWM-a}-\ref{eq:d-FWM-c}) predicts the conditions for which a $\mu$-pair of parametric fields starts to oscillate.  By calculating the  eigenvalues corresponding to Eqs.~(\ref{eq:d-FWM-a}-\ref{eq:d-FWM-c}) linearized around the cw steady state solution, one obtains~\cite{Mosca:2015wh}

\begin{equation}
\lambda_\pm = -  \gamma -  g (\eta_{\mu00\mu}+ \eta_{\bar{\mu}00\bar{\mu}}^*) |A_0|^2
\pm 
\sqrt{ g^2 |\eta_{\mu\bar{\mu}00}|^2 \, |A_0|^4
- \hspace{-3pt} \left[ \Delta_{0} - \hspace{-2pt} D_2 \mu^2 -  \hspace{-2pt} i g (\eta_{\mu00\mu}- \eta_{\bar{\mu}00\bar{\mu}}^*) |A_0|^2\right]^2 } \, ,
\label{eigen1}
\end{equation}
where $D_2 \simeq -2 \pi^2 c^3 \beta^{\prime\prime}/L^2 \, n_0^3 = -(c/2 n_0) D_1^2 \beta^{\prime\prime}$ accounts for the group velocity dispersion at $\omega_0$, with $ \beta^{\prime\prime} = \left. \frac{d^2k}{d\omega^2} \right|_{\omega_0}$, and  $n_0=n(\omega_0)$ the refractive index at $\omega_0$.
Side modes start to oscillate, i.e., the zero solution for the parametric fields becomes unstable, when the real part of an eigenvalue goes from negative to positive values. 
The coupling constants which appear in Eq.~(\ref{eigen1}) are: $\eta_{\mu\bar{\mu}00}$, which is the parametric gain related to cascaded SHG and OPO, whereby two photons at frequency $\omega_0$ annihilate and two parametric photons at $\omega_\mu$ and $\omega_{\bar{\mu}}$ are created, mediated by a SH photon; and $\eta_{\mu00\mu}$ ($\eta_{\bar{\mu}00\bar{\mu}}$), which is related to the sum frequency process between a parametric photon at $\omega_\mu$ ($\omega_{\bar{\mu}}$) and the pump. The latter process is the most relevant nonlinear loss at the threshold (second term of r.h.s of Eq.~(\ref{eigen1})), and provides a nonlinear phase shift (last term in the square brackets of r.h.s of Eq.~(\ref{eigen1})). 
The lowest threshold occurs for a pair of parametric fields which starts to grow close to the minima of the sum frequency generation (SFG) efficiency.


A general expression for the dynamic equations for any number of interacting fields can be derived heuristically~\cite{Mosca:2015wh}, yielding for each field $A_{\mu}$, nearly resonant with the  $\mu$-th cavity mode,

\begin{equation}
\dot{A}_{\mu} = -(\gamma_{\mu} + i \Delta_{\mu}) \, A_{\mu}
	- g \!\!\!\!\!\! \sum_{\substack{{\rho,\sigma}\\{\nu=\rho+\sigma-\mu}}} \!\!\!\!\!\! \eta_{\mu\nu\rho\sigma} \, A_{\nu}^* A_{\rho} A_{\sigma}  + F_\text{in} \, ,
\label{eq:FWM-gen}
\end{equation}
where the summation over the indices $\rho$ and $\sigma$ goes over all the cavity resonant modes. 
The complex coupling constants are given by Eqs.~(\ref{eq:eta}), while the constraint over $\nu$ accounts for energy conservation.
The coupled mode equations (\ref{eq:FWM-gen}) are formally analogous to the modal expansion  for Kerr combs~\cite{Chembo:2010cb,Chembo:2010ii}, and describe the whole comb dynamics.
It is worth noting that the information provided by the linear stability analysis only holds for the very beginning of comb formation. Very quickly, a large number of cavity modes under the gain curve grow from noise. At the same time, they interact with each other through multiple nonlinear processes. These processes are not considered in the linear stability analysis, which intrinsically considers only three interacting modes. The long-term spectral configuration is thus the result of a complex interaction between many modes, over thousands of cavity round-trips~\cite{Hansson:2017cs}. 
\section{Time-domain model for quadratic combs}

An alternative description of quadratic comb dynamics can be given in terms of time evolution of the slowly varying intracavity field envelopes. 
Let us define the envelopes $A(z,\tau)$ for the fundamental, and $B(z,\tau)$ for the second harmonic electric fields in a resonator.
Field dynamics can be described by an infinite dimensional map (Ikeda map) for the field amplitudes~\cite{Leo:2016kj,Leo:2016df}, which describes the evolution of cavity fields over the  $m$th round trip, along with the boundary condition for the fields at the end of each round trip.
The propagation equations for the fields $A_m(z,\tau)$ and $B_m(z,\tau)$ read as

\begin{align}
\frac{\partial A_m}{\partial z} =& \left[-\frac{\alpha_{c1}}{2}- i\frac{{k}_1''}{2}\frac{\partial^2}{\partial \tau^2}\right] \hspace{-2pt} A_m+i\kappa B_mA_m^*e^{-i \Delta k z}, 
\label{mapA}\\
\frac{\partial B_m}{\partial z} = &\left[-\frac{\alpha_{c2}}{2} - \Delta {k}'\frac{\partial }{\partial \tau}-i\frac{{k}_2''}{2}\frac{\partial^2 }{\partial \tau^2}\right] \hspace{-2pt} B_m+i\kappa A_m^2 e^{i\Delta k z} \, ,
\label{mapB}
\end{align}
where 
$z\in [0,L]$ is the position along the cavity round-trip path;
$\alpha_{c1,2}$ are propagation losses (hereafter, subscripts 1 and 2 denote fields at $\omega_0$ and $2\omega_0$, respectively); 
${k}''_{1,2} = \mathrm{d}^2k/\mathrm{d}\omega^2|_{\omega_0, 2\omega_0}$ are the group velocity dispersion coefficients; 
$\Delta {k}' = \mathrm{d}k/\mathrm{d}\omega|_{2\omega_0}-\mathrm{d}k/\mathrm{d}\omega|_{\omega_0}$ is the corresponding group-velocity mismatch, or temporal walk-off.
 The “fast-time” variable $\tau$ describes the temporal profiles of the fields in a reference frame moving with the group velocity of light at $\omega_0$. 
 
For the case of intracavity SHG, the fields at the beginning of the $(m+1)$th round trip are related to the fields at the end of the previous $m$th round trip according to the following cavity boundary conditions,

\begin{align}
A_{m+1}(0,\tau) &= \sqrt{1-\theta_1} \, A_{m}(L,\tau) \, e^{-i\delta_1} +  \sqrt{\theta_1} A_\text{in}
\label{boundA} 
\\
B_{m+1}(0,\tau) &=  \sqrt{1-\theta_2} \, B_{m}(L,\tau) \, e^{-i\delta_2},
\label{boundB} 
\end{align}
where $\theta_{1,2}$ are power transmission coefficients at the coupling mirror, $\delta_{1}\simeq(\omega_{0}-\omega_{c1})t_\text{R}$ and $\delta_{2}\simeq(2\omega_{0}-\omega_{c2})t_\text{R}$ are the round-trip phase detunings for the fields at $\omega_0$ and $2\omega_0$, respectively, with  $\omega_{c1}$ and $\omega_{c2}$ the frequencies of the respective nearest cavity resonance, and $A_\text{in}$ is the external, constant  driving field amplitude.
It is worth noting that the Ikeda map of Eqs.~(\ref{mapA})-(\ref{boundB}) can describe different nonlinear systems (SHG or OPO, either singly or doubly resonant), by suitably choosing the boundary conditions. 
For a singly resonant cavity SHG, $\theta_2=1$, and the SH field resets at the beginning of each round trip, i.e., $B_{m+1}(0,\tau) =  0$. 

\begin{figure}[t]
\centering
\includegraphics*[viewport= 0 0 560 280, clip,width=10 cm]{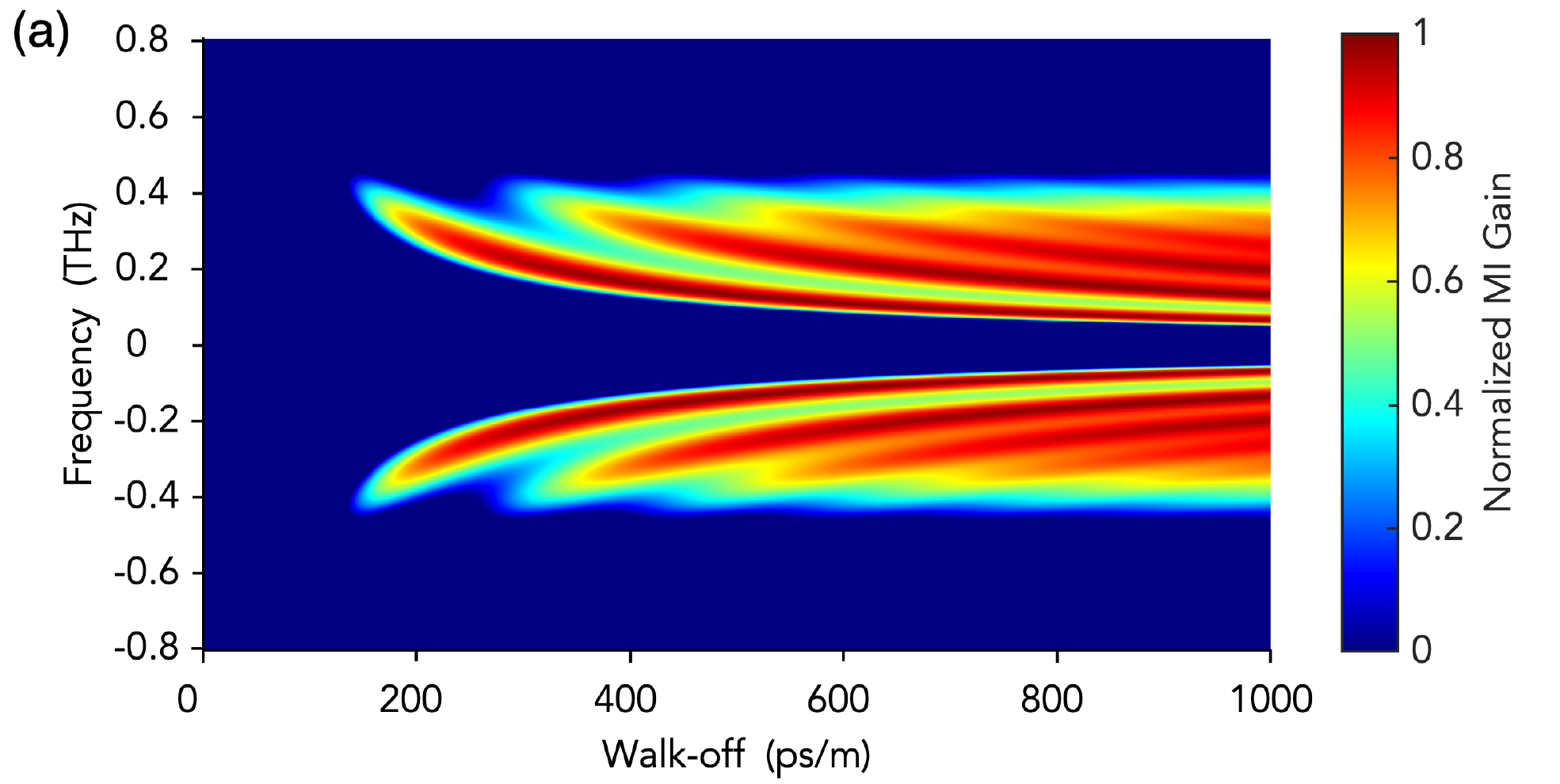}
\includegraphics*[viewport= 10 0 560 280, clip,width=10 cm]{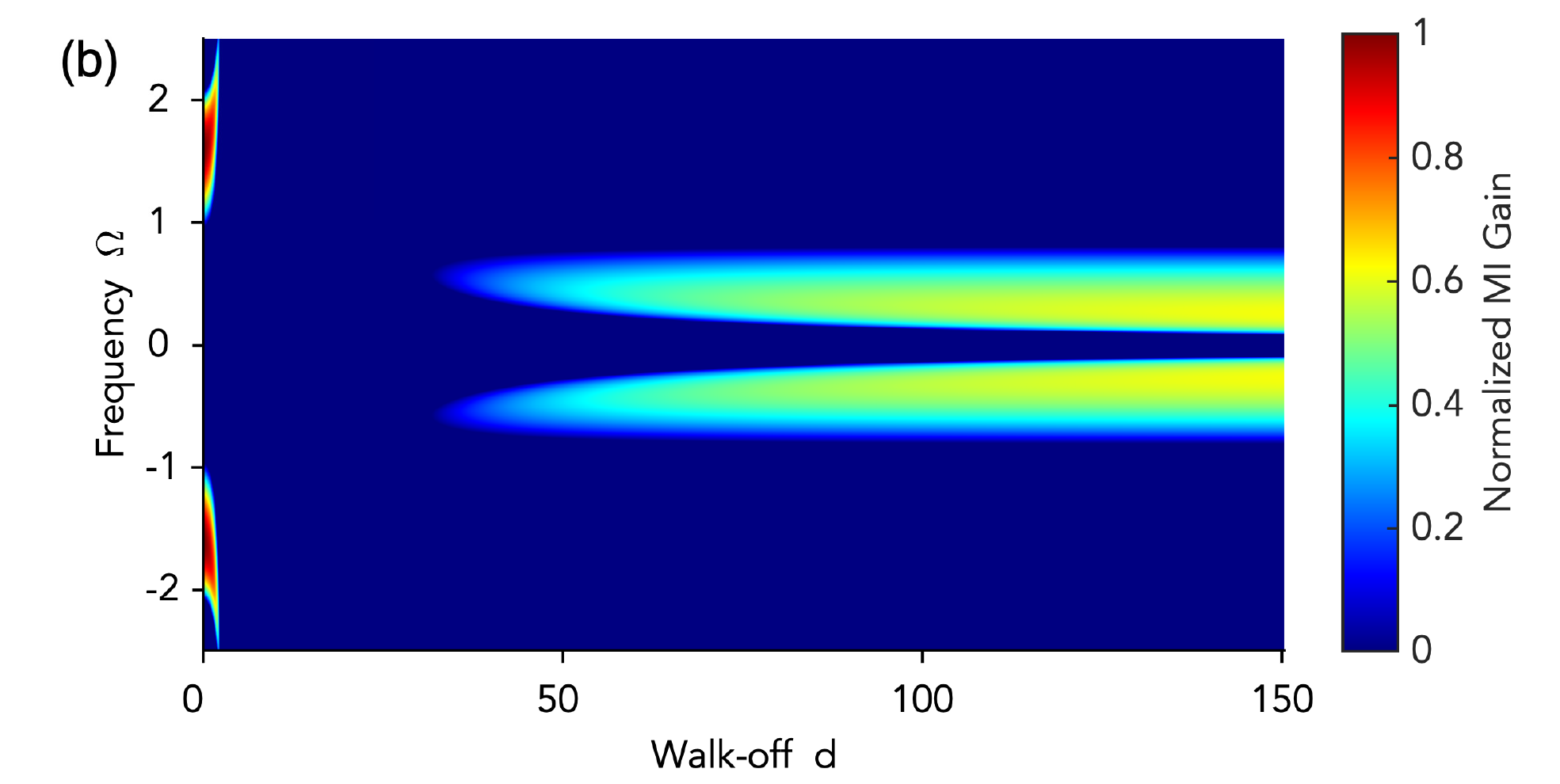}
\caption{Modulation instability gain profiles as a function of temporal walk-off. (a) Singly resonant cavity SHG. (b) Doubly resonant cavity SHG (parameters are normalized according to Ref.~\cite{Leo:2016df}). Adapted with permission from ~\cite{Leo:2016kj,Leo:2016df}. Copyrighted by the American Physical Society.}
\label{fig:MIgain}
\end{figure}   

For a relatively high-finesse resonator, the fundamental field evolves slowly during each round trip, and the infinite dimensional map may be averaged over one round trip length $L$. This averaging procedure yields a single mean field equation for the fundamental field amplitude~\cite{Leo:2016kj},
 
\begin{align}
 t_\mathrm{R}\frac{\partial A(t,\tau)}{\partial t} =& \bigg[ -\alpha_1 - i \delta_1 -iL\frac{{k}_1''}{2}\frac{\partial^2}{\partial \tau^2} \bigg]A 
 - \rho A^*\left[A^2(t,\tau)\otimes I(\tau)\right] + \sqrt{\theta_1}\,A_\mathrm{in} \, ,
   \label{MF}
\end{align}
where $t$ is a ``slow time'' variable, linked to the roundtrip index as $A(t=mt_\mathrm{R},\tau) = A_m(z=0,\tau)$~\cite{Haelterman:1992cd,Leo:2010if,Coen:2013hw,Coen:2013hd}, $\alpha_1 = (\alpha_{c1}L+\theta_1)/2$, $\rho = (\kappa L)^2$, $\otimes$ denotes convolution and the nonlinear response function $I(\tau) = \mathscr{F}^{-1}[\hat{I}(\Omega)]$, with $\hat{I}(\Omega) = \left[(1-e^{-ix}-ix)/x^2\right]$, $x(\Omega) = \left[\Delta k+i\hat{k}(\Omega)\right]L$, and
$\hat{k}(\Omega) = -\alpha_{c,2}/2+i\left[\Delta{k}'\Omega+({k}_2''/2)\Omega^2\right]$.
Here, we define the direct and inverse Fourier transform operator as $\mathscr{F}\left[\cdot\right]=\int_{-\infty}^{\infty}\cdot\,e^{i\Omega\tau}\,\mathrm{d}\tau$ and $\mathscr{F}^{-1}\left[\cdot\right]= (2\pi)^{-1}\int_{-\infty}^{\infty}\cdot\,e^{-i\Omega\tau}\,\mathrm{d}\Omega$, respectively.

Similarly to the coupled mode equations in frequency domain, also the mean field equation (\ref{MF}) exhibits an effective cubic nonlinearity, with a noninstantaneous response analogous to the delayed Raman response of cubic nonlinear media and other generalized nonlinear Schr\"odinger models. 

Linear stability analysis of the cw solution of Eq.~(\ref{MF}) leads to the following expression for the eigenvalues~\cite{Leo:2016kj},

\begin{equation}
\lambda_\pm = \hspace{-2pt} -\left(\alpha_1+\rho P_0[\hat{I}(\Omega)+\hat{I}^*(-\Omega)]\right) \pm \sqrt{|\hat{I}(0)|^2\rho^2P_0^2 - \hspace{-2pt} \left(\delta_1- \hspace{-2pt} \frac{{k}_1''L}{2}\Omega^2-i\rho P_0[\hat{I}(\Omega)-\hat{I}^*(-\Omega)]\right)^2} \, ,
\label{MIgain}
\end{equation}
which, baring the notation, is substantially equivalent to Eq.~(\ref{eigen1}).
Figure \ref{fig:MIgain}(a) shows the MI gain, $\Re[\lambda_+]$ profile as a function of the walk-off parameter $\Delta k'$. Clearly, there is no MI for zero walk-off, and MI appears for sufficiently large values of walk-off, revealing the fundamental role of group-velocity mismatch for the formation of quadratic optical frequency combs and related dissipative temporal patterns.  

\begin{figure}[t]
\centering
\includegraphics*[viewport= 0 280 370 470, clip,width=16 cm]{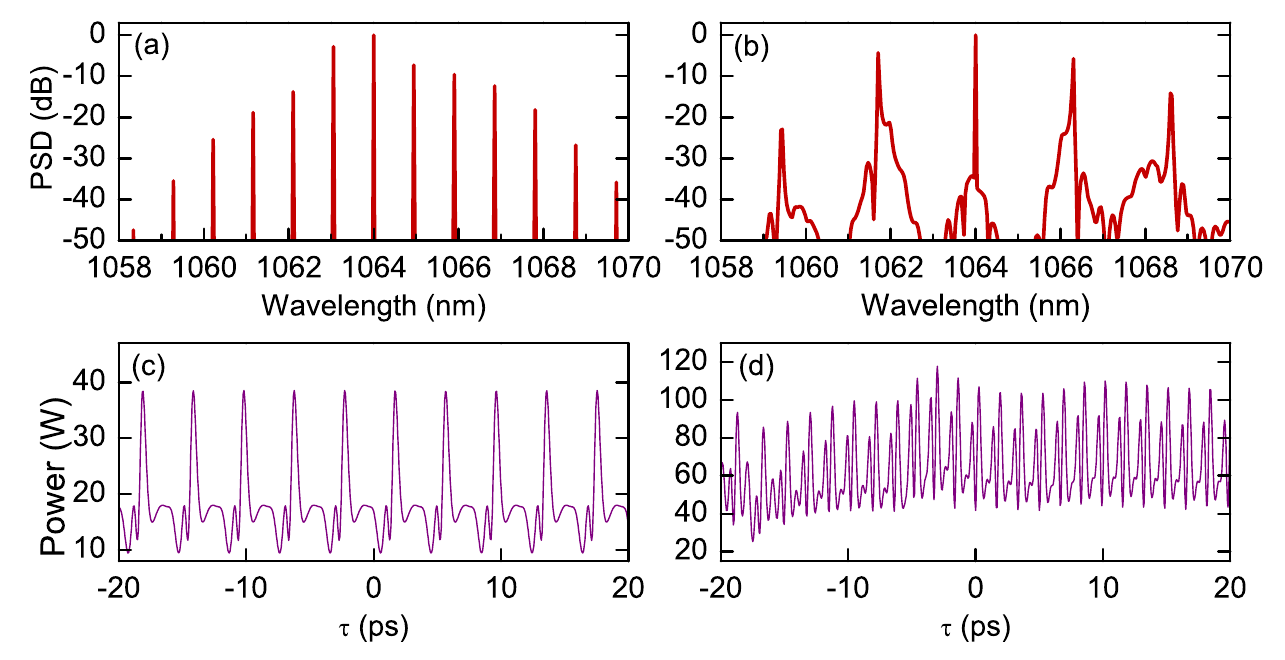}
\caption{Numerical simulation of Eq.~(\ref{MF}), using the parameters of the system in Ref.~\cite{Ricciardi:2015bw}. (a) Input power 2~W, $\delta_1=0.001$. (b) Input power 7~W, $\delta_1=0.01$. (c) and (d) Details of the respective temporal patterns.  }
\label{fig:SimSR}
\end{figure}

Hansson et al.~\cite{Hansson:2017cs} demonstrated that the general system of coupled mode equations (\ref{eq:FWM-gen}) can be derived from the map of Eqs.~(\ref{mapA})-(\ref{boundB}). However, frequency domain coupled mode equations are not exactly equivalent to the time domain mean field Eq.~(\ref{MF}): the two approaches differ in the way the dispersion is averaged, although they provide almost equal results for the system of Ref.~\cite{Ricciardi:2015bw}. 

Theoretical models, in addition to providing useful insight into the physics of quadratic combs, can be a practical tool for simulating the comb dynamics, giving access to information not always available from the experiment.
Both the frequency and time domain formalisms here described lend themselves to the numerical simulation of comb dynamics. 
Coupled mode equations (\ref{eq:FWM-gen}) are in general more time consuming than time domain approaches, unless they can be cast in a way where fast Fourier transform (FFT) algorithms can effectively reduce the computation time~\cite{Hansson:2014ie}.
Numerical integration of the Ikeda map or the derived mean-field equation usually relies on split-step Fourier methods~\cite{Agrawal:2001book,Weideman:1986hc}. 
According to this methods, propagation along each integration step is carried out in two steps. In a first step, the nonlinear and driving terms are propagated by means of a 4th-order Runge--Kutta method. The dispersive and absorption terms are propagated in a second step, where their propagation operator is evaluated in the Fourier domain, using a FFT algorithm.
The simulation initiates by assuming a constant amplitude, input driving field $A_\text{in}$ that describes the resonant pump laser. More importantly, in the first step a numerical white-noise background of one photon per mode must be added in order to seed the nonlinear processes which lead to the comb.
Whereas the numerical integration of Ikeda map requires a spatial step size smaller than the cavity round trip length, the mean-field equation can be numerically integrated with temporal step sizes of the order of the round-trip time, for the benefit of the computation time.

Figure \ref{fig:SimSR} shows two spectra, (a) and (b), and the respective temporal patterns, (c) and (d), obtained by numerically integrating Eq.~(\ref{MF}). 
The simulations have been performed using the parameters from Ref.~\cite{Ricciardi:2015bw}, in the case of quasi-phase matched SHG, for a constant input power of 2 and 7~W, respectively, and a small positive detuning.  
The simulated spectra are in good agreement with the experimental spectra shown in Fig.~\ref{fig:SR-PM}(b) and (c). For the moment, we cannot determine the temporal profile corresponding to a comb spectra. Hence, numerical simulations provide insights on the temporal feature of comb dynamics.
We notice that the temporal pattern (c) associated to spectrum (a) has a stable periodic structure (also called Turing or roll pattern), which entails a strong phase coupling between the spectral modes, i.e., a mode-locked regime. Instead, the spectrum of Fig.~\ref{fig:SimSR}(b), with secondary combs around the primary sidebands, corresponds to an irregular temporal pattern with no evidence of intermodal phase coupling. Moreover, it does not appear to reach a stationary regime. 
In both cases, the emission is not purely pulsed, as typically occurs for combs generated in  femtosecond, mode-locked lasers, but the temporal patterns coexist with a flat background. 
The coexistence of a temporal pattern with a flat background is frequent for Kerr combs~\cite{Godey:2014ks}, as well as for combs generated in quantum cascade lasers\cite{Khurgin:2014hy,Cappelli:2019bn}.  
In fact, in femtosecond laser combs the emission of short pulses is due to a particular phase relation between laser mode---i.e., all the modes have equal phases. 
However, in a wider sense, mode-locking only requires that a stable phase relation holds between all the mode fields.
Finally, numerical simulations also reveal a slow drift of the temporal patterns (both at the fundamental and the SH fields) in the reference frame moving with the group velocity of the FF.


When $\theta_2<1$, the infinite dimensional map of Eqs.~(\ref{mapA})-(\ref{boundB}) describes the case of a doubly resonant optical cavity, where also second harmonic fields may resonate. Leo et al. theoretically analyzed this system~\cite{Leo:2016df} and derived a couple of two mean-field equations, which accurately model comb dynamics. These equations read, assuming phase-matched SHG,

 \begin{align}
t_\mathrm{R}\frac{\partial A}{\partial t} &= \left[ -\alpha_1-i\delta_1- i\frac{{k}_1''L}{2}\frac{\partial^2}{\partial \tau^2} \right] \hspace{-2pt} A +i\kappa L BA^* + \sqrt{\theta_1}A_\mathrm{in}, 
\label{MF_F} \\
t_\mathrm{R}\frac{\partial B}{\partial t} &= \left[ -\alpha_2-i\delta_2- \Delta {k}'L\frac{\partial }{\partial \tau}- i\frac{{k}_2''L}{2}\frac{\partial^2}{\partial \tau^2} \right] \hspace{-2pt}  B + i\kappa L A^2 \, ,
\label{MF_S}
\end{align} 
where $\alpha_2$ is the cavity loss of the SH field.

Under realistic conditions, the two mean-field equations (\ref{MF_F}) and (\ref{MF_S}) can be reduced to a single mean-field equation, analogously to Eq.~(\ref{MF}) for singly resonant cavity SHG. One obtains 

\begin{equation}
t_\mathrm{R}\frac{\partial A}{\partial t} = \left[ -\alpha_1-i\delta_1- i\frac{{k}_1''L}{2}\frac{\partial^2}{\partial \tau^2} \right]A 
-\rho  A^* \left[A^2 \otimes J\right] + \sqrt{\theta_1}A_\mathrm{in},
\label{MF-DR}
\end{equation}
where the Fourier transform of the kernel function $J$  is 
\begin{equation}
\hat{J}(\Omega) = \frac{1}{\alpha_2+i\delta_2 - i\Delta {k}'L\Omega - i\frac{{k}_2''L}{2}\Omega^2}.
\end{equation}
A linear stability analysis of the cw solution (for both the Ikeda map and the mean-field approximations) reveals the significant role of temporal walk-off in enabling comb formation. However, in this case, MI gain may also occur for zero or relatively small values of the walk-off [Fig.~\ref{fig:MIgain}(b)].

\begin{figure}[t]
\centering
\includegraphics*[viewport= 0 0 560 280, clip,width=10 cm]{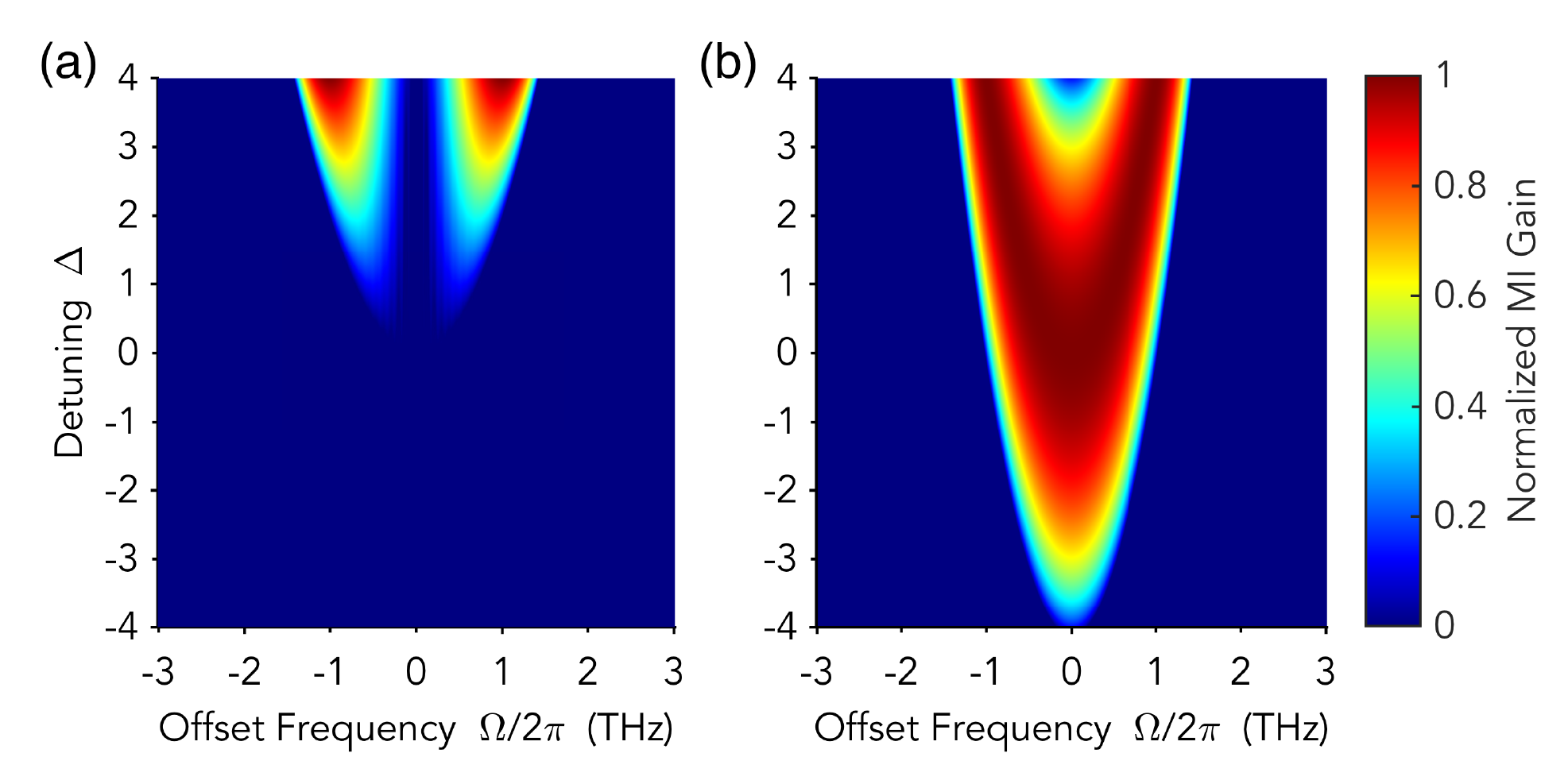}
\includegraphics*[viewport= -20 0 560 320, clip,width=10 cm]{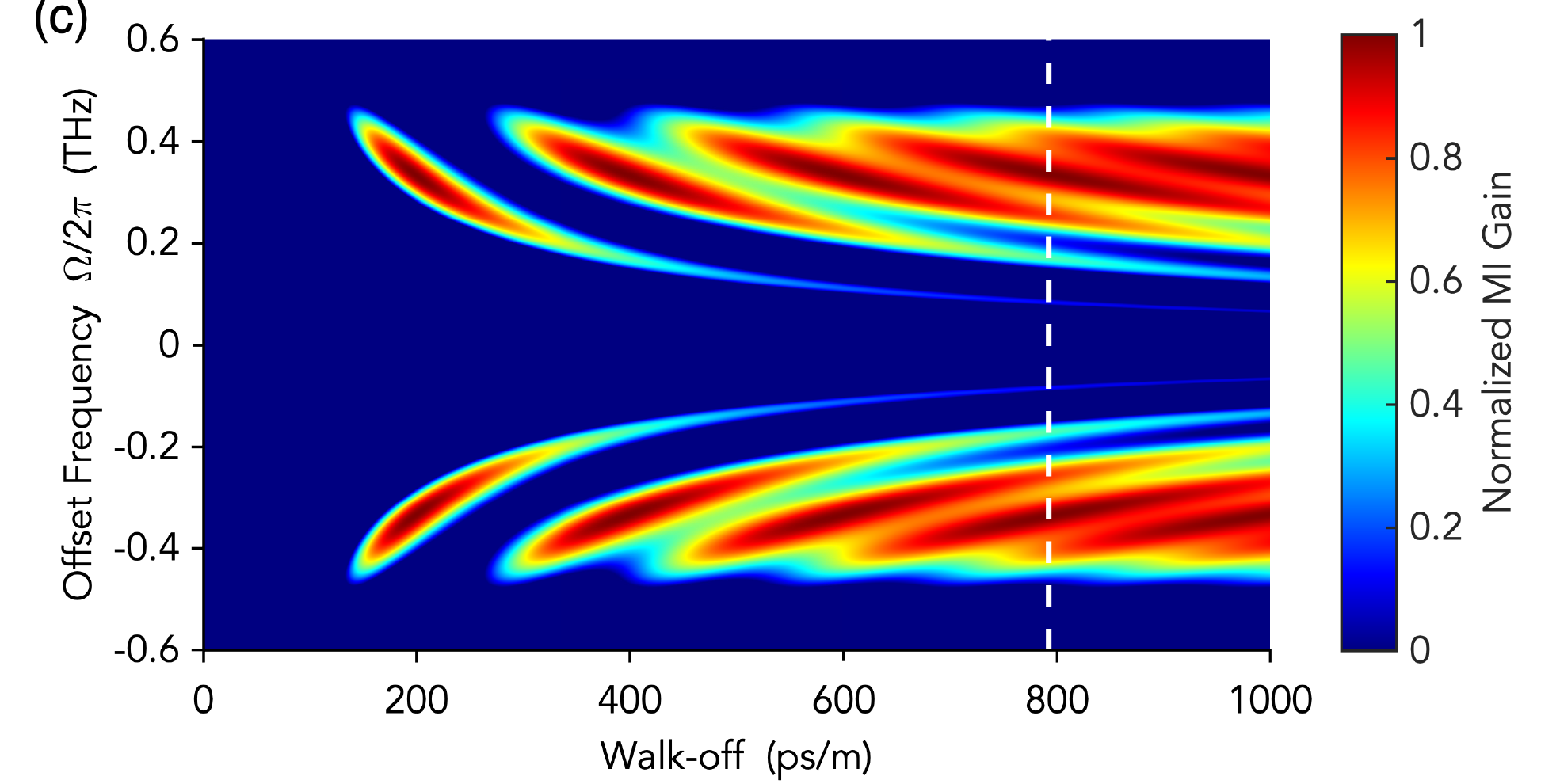}
\caption{OFC in a degenerate OPO. (a) and (b) show the MI gain as a function of the normalized cavity detuning $\Delta=\delta_1/\alpha_1$, for the constant solution and the zero solution, respectively. (c) MI gain profiles as a function of the temporal walk-off. Adapted  with permission from~\cite{Mosca:2018jk}. Copyrighted by the American Physical Society.}
\label{fig:MI_OPO}
\end{figure}

\section{Combs in optical parametric oscillators}
Degenerate optical parametric oscillation is the inverse process of cavity SHG, when the pump field $A_\text{in}$ at the FF $\omega_0$ is replaced by a pump field $B_\text{in}$ at the SH frequency $2\omega_0$. 
Its dynamics can be described by an infinite dimensional map as well, where, in addition to Eqs.~(\ref{mapA}) and (\ref{mapB}), the following boundary conditions hold for the fields at the beginning of each round trip,

\begin{align}
A_{m+1}(0,\tau) &= \sqrt{1-\theta_1} \, A_{m}(L,\tau) \, e^{-i\delta_1}
\label{OPOboundA} 
\\
B_{m+1}(0,\tau) &= B_\mathrm{in} \,  .
\label{OPOboundB} 
\end{align}
Here, we consider an OPO cavity where only the parametric field resonates. It is straightforward to extend the analysis to the case when also the harmonic pump field resonates.

\begin{figure}[t!]
\centering
\includegraphics*[viewport= 0 0 540 420, clip,width=14 cm]{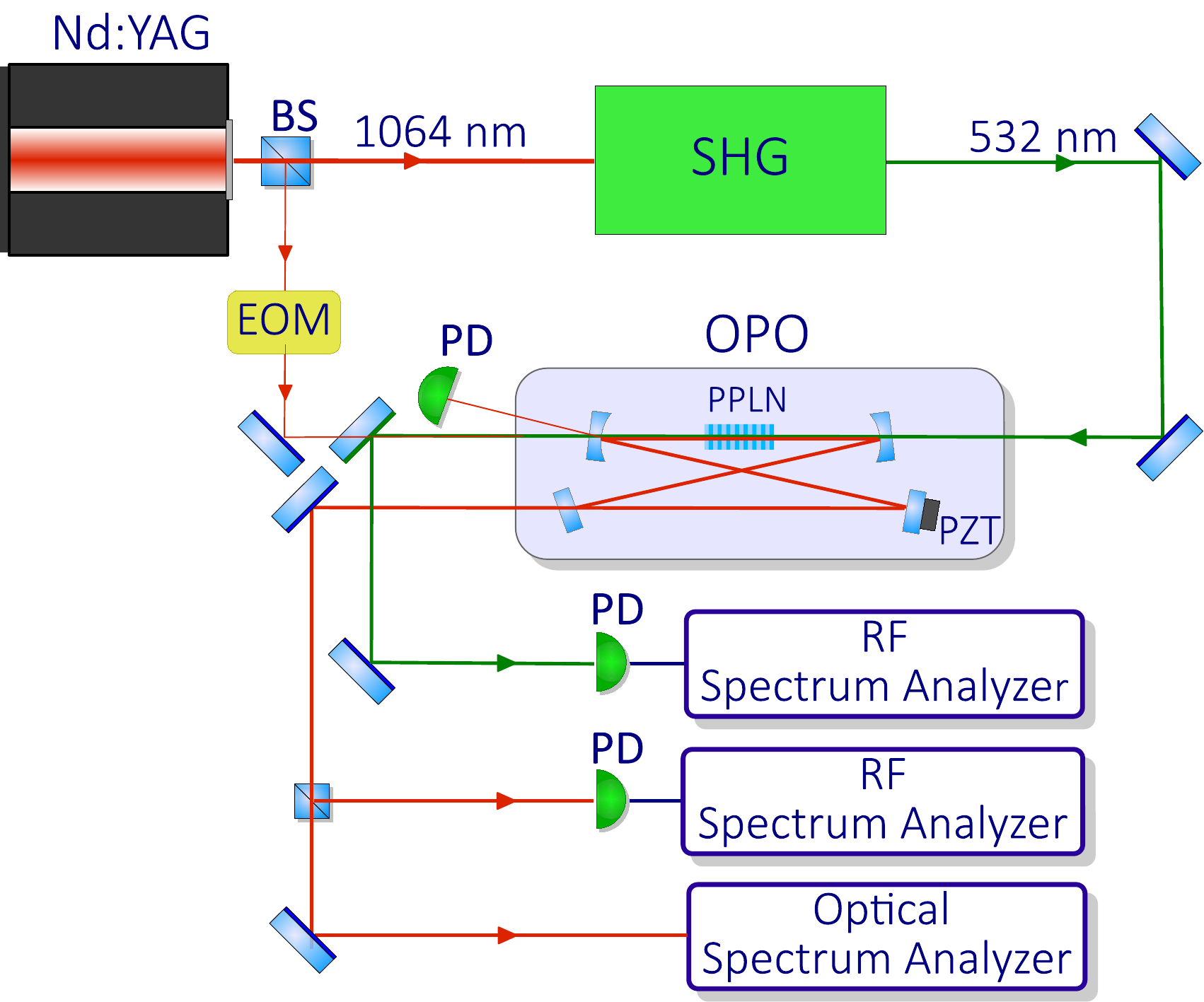}
\caption{OFC in a degenerate OPO. Scheme of the experimental setup: beam splitter (BS), electro-optic phase modulator (EOM), periodically poled lithium niobate crystal (PPLN), piezoelectric actuator (PZT), photodiode (PD). Adapted  with permission from~\cite{Mosca:2018jk}. Copyrighted by the American Physical Society.}
\label{fig:OPO}
\end{figure}

Following the approach of Ref.~\cite{Leo:2016kj}, the infinite dimensional map can be combined into a single mean-field equation for the parametric field $A$, which reads, assuming $\Delta k =0$~\cite{Mosca:2018jk},

\begin{align}
  t_\mathrm{R}& \frac{\partial A(t,\tau)}{\partial t}  = 
   \left[ -\alpha_1 - i \delta_1 -i\frac{L{k}_1''}{2}\frac{\partial^2}{\partial \tau^2} \right] \, A (t,\tau)
   - \mu^2 A^*(t,\tau)\left[A^2(t,\tau)\otimes I(\tau)\right] 
   + i \mu  B_\mathrm{in} A^*(t,\tau) \, ,
 \label{MFE}
\end{align}
where all the physical parameters and the kernel function $I$ are the same as in Eq.~(\ref{MF}).
We note that Eq.~(\ref{MFE}) is similar to the corresponding mean-field equation for comb dynamics in cavity SHG, except for the parametric driving force (last term on the r.h.s.).
Equation (\ref{MFE}) has a trivial zero solution, $A_0=0$,  and a nontrivial time independent solution, $A_0=|A_0|e^{i\phi}$.  
From a  linear stability analysis of the constant solution, we derived the following expression for the eigenvalues~\cite{Mosca:2018jk},

\begin{equation}
\lambda_{\pm} =  
- \left[ \alpha_1 + \mu^2 |A_0|^2 {\cal I}_+(\Omega) \right] 
 \pm
\sqrt{ 
(\alpha_1^2 + \delta_1^2) \hspace{-2pt}
- \hspace{-2pt} \left[ \delta_1 \hspace{-2pt} - \hspace{-2pt}D_2  \Omega^2  - i \mu^2 |A_0|^2 \, {\cal I}_-(\Omega)  \right]^2
} \, ,
\label{auto1}
\end{equation}
where $|A_0|^2=[-\alpha_1 \pm \sqrt{\mu^2 B_\mathrm{in}^2 -\delta_1^2}]  /\mu^2  \, \hat{I}(0)$ is the squared modulus of the nontrivial solution and ${\cal I}_\pm(\Omega) = \hat{I}(\Omega) \pm \hat{I}^*(-\Omega)$. 
Similarly, for the zero solution the eigenvalues of the linearized system are

\begin{align}
\lambda_{\pm} =&  
-  \alpha_1 \pm
\sqrt{ \mu^2 B_\mathrm{in}^2
- \left( \delta_1 - D_2  \Omega^2 \right)^2} \, .
\label{auto0}
\end{align}
Both solutions exhibit MI gain for $\mathrm{Re}[\lambda_+]>0$, which are shown in Fig.~\ref{fig:MI_OPO}(a) and (b) as a function of the cavity detuning.
From Eqs.~(\ref{auto1}) clearly appears that MI gain for the nontrivial solution depends both on walk-off $\Delta k'$, through ${\cal I}_\pm(\Omega)$, and GVD.
As for singly resonant cavity SHG, MI only manifests itself for relatively high walk-off values, while it is absent for zero walk-off, as shown in Fig.~\ref{fig:MI_OPO}(c).  
The instability of the zero solution, which is not expected in the usual dispersionless analysis of the OPO, does not depend on the walk-off, but it is rather induced by GVD. 
Actually, GVD is responsible for the unequal spacing between cavity resonances, so that they are asymmetrically displaced with respect to the degeneracy frequency $\omega_0$, when the latter is perfectly resonant. Thus, GVD effectively favors parametric oscillations close to the degeneracy frequency.
For normal dispersion, a positive detuning between the degeneracy frequency and the nearest cavity resonance can make symmetric an initially asymmetric pair of distant resonances, which now can more favourably oscillate than the degeneracy frequency $\omega_0$. 
The larger the detuning, the more distant the symmetric resonances are. 
For small negative detunings no resonance pair can be symmetrically displaced around $\omega_0$, and MI gain is maximum at the degeneracy frequency, decreasing as a function of the detuning amplitude.
The same occurs in the case of anomalous dispersion, provided that the detuning sign is reversed.

\begin{figure}[t!]
\centering
\includegraphics*[viewport= 0 0 680 430, clip, width=15 cm]{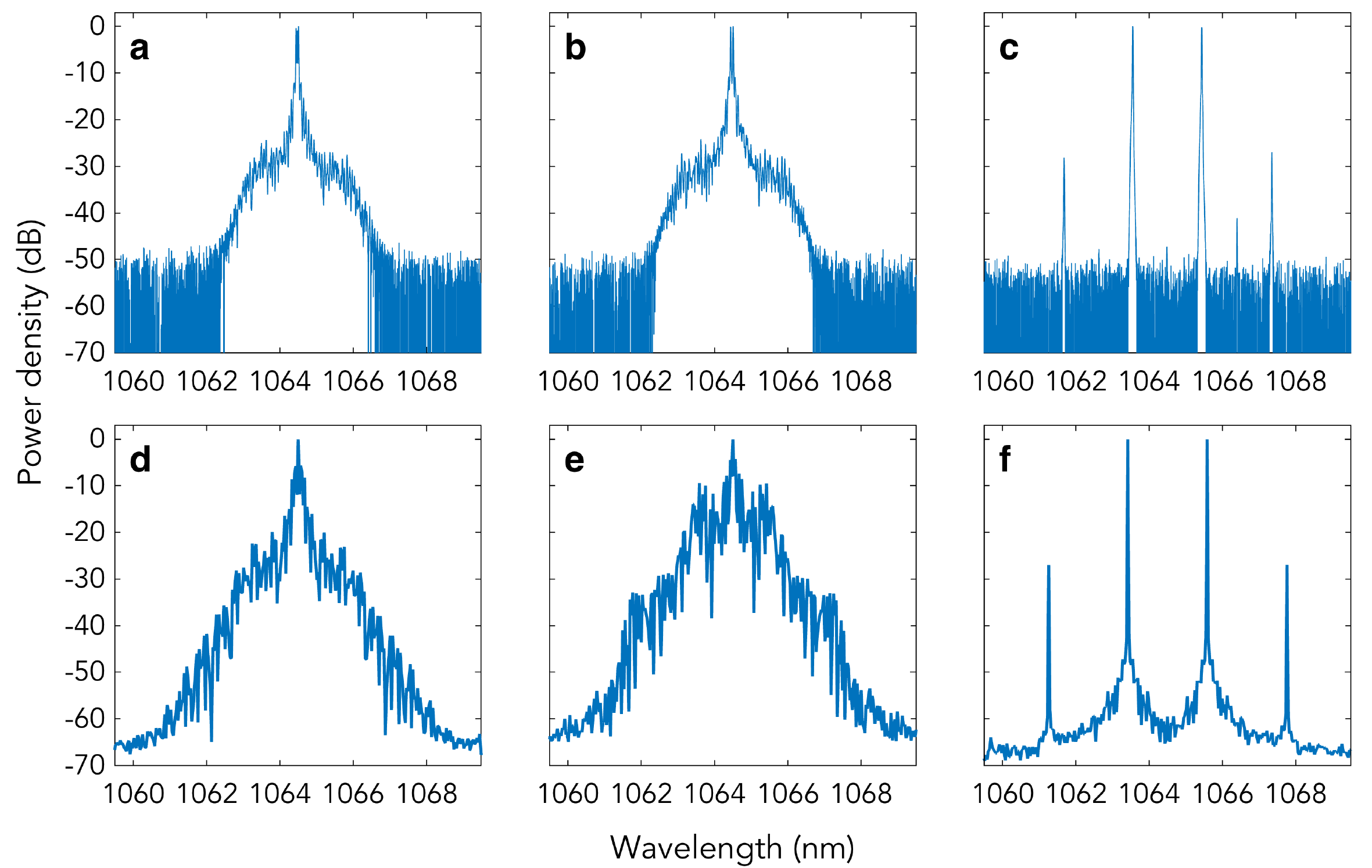}
\caption{(a)-(c) Experimental OPO optical spectra for detunings $\Delta=-0.30, 0.00, 0.30$, respectively. (d)-(f) Corresponding numerically calculated spectra. From~\cite{Mosca:2018jk}.  Copyrighted by the American Physical Society.}
\label{fig:spectra6}
\end{figure}

Frequency comb generation in an OPO has been demonstrated by using a nearly degenerate OPO pumped by a frequency doubled cw Nd:YAG laser (Fig.~\ref{fig:OPO}). The OPO was based on a 15-mm-long periodically-poled 5\%-MgO-doped lithium niobate crystal, with a grating period of $\Lambda = 6.92\;\mu \text{m}$, enclosed in a bow-tie cavity resonating for the parametric wavelengths around 1064 nm, similar to that used for cavity SHG.  The nonlinear crystal was located between two high-reflectivity spherical mirrors (with radius of curvature = 100 mm), while  a flat high-reflectivity mirror was mounted on a piezoelectric actuator for cavity length control. A fourth, partially reflective flat mirror (R = 98\%) allowed us to couple out the generated parametric radiation. The SH beam entered the OPO cavity from a first spherical mirror, passed through the nonlinear crystal, and left the cavity at the second spherical mirror. The FSR of the cavity was $505$~MHz.
We observed combs for pump powers higher than 85~mW (about three times the OPO threshold of 30~mW), and studied the effect of small cavity detunings on the comb spectra.
Figures~\ref{fig:spectra6}(a)–\ref{fig:spectra6}(c) show the experimental comb spectra recorded for $\Delta=-0.30, 0.00, 0.30$, respectively, with 300~mW of pump power. 
We found a good agreement with the corresponding spectra, shown in Figs.~\ref{fig:spectra6}(d)–\ref{fig:spectra6}(f), calculated by numerically integrating the mean-field equation (\ref{MFE}).   
Experimental spectra for negative and zero detunings are very similar, displaying  1 FSR line spacing, whereas for the positive detuning the experimental spectrum consists of two pairs of widely spaced symmetric lines.

\section{Single envelope equation}
Models based on the two field envelopes, i.e., Eqs.~(\ref{mapA})-(\ref{boundB}), and their approximations hold as long as there is a single dominant nonlinear process, and the combs are confined around two carrier frequencies. When the combs start to overlap, or multiple nonlinear processes play a prominent role, frequency comb generation may be studied by means of a more general model, based on a single-envelope equation combined with the  boundary conditions that relate the fields between successive round trips and the input pump field~\cite{Hansson:2016kz}, 
  
\begin{equation}
\mathcal{F} [A^{m+1} (\tau,0)] = \sqrt{ \hat{\theta}(\Omega)}  \mathcal{F}[A_\text{in}] + \sqrt{1 -\hat{\theta}(\Omega)} \, e^{i \phi_0}  \mathcal{F} [A^{m} (\tau,L)] 
\label{SEE1}
\end{equation}

\begin{equation}
\left[ \partial_z -D\left( i \frac{\partial}{\partial \tau}\right) + \frac{\alpha_d}{2} \right] A^{m} (\tau,z) = 
i \rho_0 \left( 1 + i\tau_\text{sh} \frac{\partial}{\partial \tau}  \right) p_\text{NL} (\tau,z,A^{m}) \, .
\label{SEE2}
\end{equation}

The boundary condition, Eq.~(\ref{SEE1}), is written in the Fourier domain, in order to account for the frequency dependence of the transmission coefficient $\theta$ at the input port of the resonator.
It determines the intra-cavity field $A_{m+1}(\tau, z = 0)$ at the beginning of $(m+1)$th round trip  in terms of the field at the end of the previous round trip $A_m(\tau, z = L)$ and the pump field $A_\text{in}$.
Equation (\ref{SEE2}) is written in a reference frame moving at the group velocity at $\omega_0$: $p_\text{NL}$  is the broadband envelope of the nonlinear polarization $P_\text{NL} = P^{(2)}_\text{NL} + P^{(3)}_\text{NL} +...  = \epsilon ( \chi^{(2)} E^2 +  \chi^{(3)} E^3 +...)$; $\rho_0=\omega_0/2n_0c\epsilon_0$; $\tau_\text{sh} =1/\omega_0$ is the shock coefficient that describes the frequency dependence of the nonlinearity; and $\alpha_d$ is the distributed linear loss coefficient.
Dispersion to all orders is included by the operator $D$,

\begin{equation}
D \hspace{-3pt}\left( i \frac{\partial}{\partial \tau}\right) = \sum_{l \ge 2} i \frac{\beta_l}{l!}  \left( i \frac{\partial}{\partial \tau}\right)^{\hspace{-3pt} l} \, ,
\end{equation}
where $\beta_l=(d^l\beta/d\omega^l)_{\omega=\omega_0}$ are expansion coefficients of the propagation constant $\beta(\omega)$.
\begin{figure}[t]
\centering
\includegraphics*[viewport= 10 0 870 300, clip,width=15 cm]{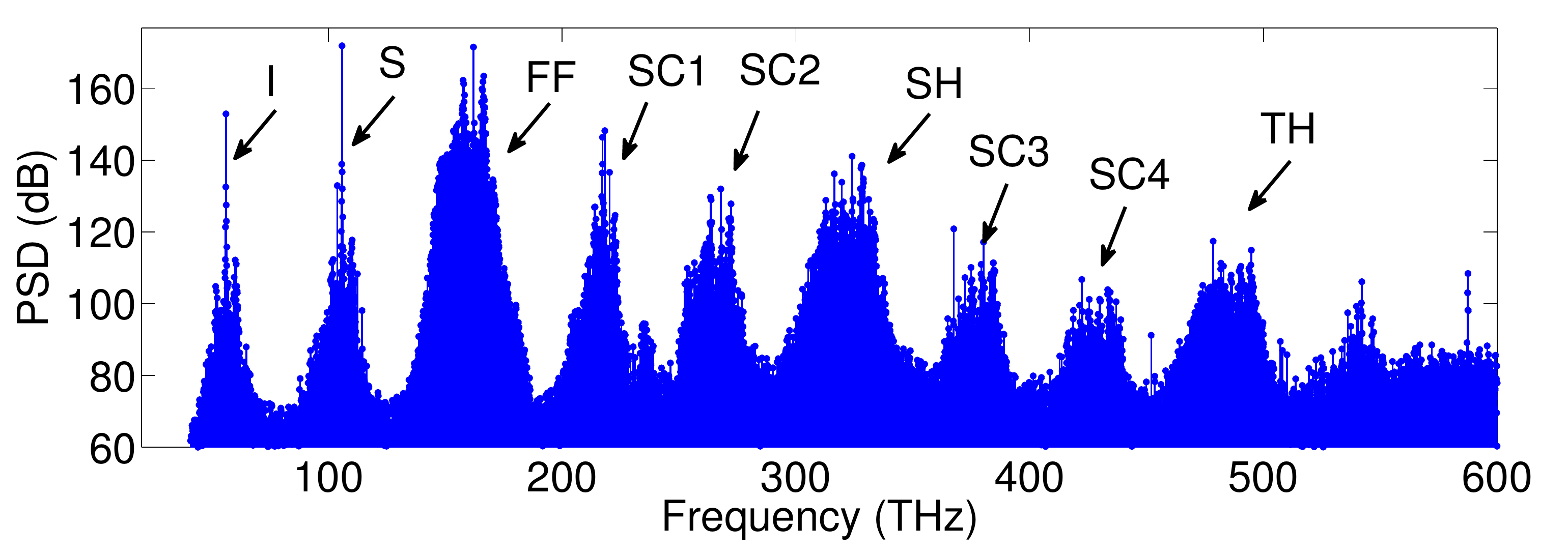}
\caption{Numerical simulation of the single-envelope map when SHG and OPO processes are simultaneously phase matched in a lithium niobate microresonator pumped by 100~mW of cw power at 1850~nm. Reprinted with permission from~\cite{Hansson:2016kz}  \copyright\ The Optical Society.}
\label{fig:SEE_OPO}
\end{figure} 

Figure \ref{fig:SEE_OPO} shows a spectrum obtained from the numerical simulation of Eqs.~(\ref{SEE1}) and (\ref{SEE2}), when SHG and non-degenerate optical parametric oscillation are simultaneously quasi-phase matched in a radially poled, lithium niobate microresonator, pumped at 1850~nm (162~THz). In this case, the quasi-phase matching period for SHG (25.56~$\mu$m) simultaneously quasi-phase matches a nondegenerate OPO with idler (signal) at 56~THz (106~THz). 
The broadband power spectral density shows a generation of a multi-comb array, extending from the mid-infrared into the ultraviolet with a spacing of a single FSR (around 92~GHz). In addition to combs at the FF, SH, and third-harmonic (TH), two additional combs are generated around signal and idler frequencies. Moreover, several secondary combs appear between the FF and the SH, and between the SH and the TH, respectively.
 These combs are generated by sum-frequency generation and difference frequency generation processes. 
For instance, the comb SC1 centered at 218~THz results from SFG between the idler and the FF, while SC3 (around 380~THz) results from SFG between the idler and the SH. 
On the other hand, DFG between the SH (TH) and the idler leads to a secondary comb SC2 (SC4) centered at 268~THz (430~THz).

\section{Perspectives}
Because of the intrinsically higher strength of the quadratic nonlinearity with respect to the third-order one, quadratic comb generation can be less demanding in terms of power density and cavity quality factor. 
Although quadratic combs have been generated in bulk cavities with moderate pump powers, their performance could increase if implemented in miniaturized devices, thus further extending and stimulating new applications~\cite{Breunig:2016ku,Strekalov:2016iw}. 
Scaling the resonator to micrometric size may be beneficial for quadratic combs, allowing for a dramatic reduction of threshold power and a flexible management of the dispersion through a geometric design, allowing for a broader comb emission.
As a matter of fact, direct generation of quadratic frequency combs has been very recently observed in chip-scale lithium niobate devices, such as periodically poled linear waveguide resonators~\cite{Ikuta:2018iw,Stefszky:2018bd}, or exploiting naturally phase-matched SHG in whispering-gallery-mode resonators~\cite{Hendry:2019tb,Szabados:2019tf}.

Several materials with second-order nonlinearity are suitable to be shaped into low-loss small-footprint resonators. 
Most of them have been used to generate Kerr combs~\cite{Levy:2011wp,Jung:2014jt,Pu:2016en,Wang:2019hn,Zhang:2019fe}, and, in some cases, secondary quadratic effects have been reported~\cite{Levy:2011wp,Jung:2014jt} or explicitly considered~\cite{Xue:2016ja}.
In contrast to Kerr combs, quadratic combs usually require more stringent conditions on phase matching and group velocity mismatch between different spectral components.
Natural~\cite{Furst:2010bt,Furst:2010co}, cyclic~\cite{Lin:2013fx}, and quasi-\cite{Meisenheimer:2015dn,Mohageg:2005cx} phase matching have been used in crystalline whispering-gallery-mode resonators.
More recently, significant progress has been made in the fabrication of integrated, high-Q, lithium niobate microresonators for $\chi^{(2)}$ processes~\cite{Guarino:2007bv,Wang:2014bq,Liang:2017bz,Wu:2018bb}.
III-V materials provide an interesting photonic platform for second-order nonlinear optics, and different techniques have been devised to achieve phase matching~\cite{Helmy:2010jy}, in particular for resonant structures~\cite{Kuo:2014aa,Mariani:2014gw,Parisi:2017fp}. 
It is worth noting that also the well developed silicon platform can be exploited for second-order nonlinear interaction. In fact, Timurdogan et al. demonstrated that a large ``dressed'' $\chi^{(2)}$ nonlinearity can be induced by breaking the crystalline center-symmetry of silicon when a direct-current field is applied across p-i-n junctions in ridge waveguides~\cite{Timurdogan:2017jg}, enabling the implementation of quasi-phase matching schemes. 

Unlike optical frequency combs in mode-locked lasers, parametrically generated combs do not usually correspond to a stable pulsed emission in the time domain. Different temporal regimes are possible, from chaotic to perfectly coherent states.
The formation of temporal cavity solitons in a cw-pumped nonlinear resonator has attracted a particular interest in connection with parametrically generated combs~\cite{Herr:2014ip}. 
Combs associated to a cavity soliton are broadband and highly coherent, which makes them ideal for low noise and metrological applications.
In fact, cavity solitons are  robust states which circulate indefinitely in a cavity, thanks to the double compensation between nonlinearity and chromatic dispersion, and between cavity losses and cw driving.
Recent theoretical works aim at identifying the dynamical regimes that exhibit soliton states  or localized solutions in cavity SHG systems~\cite{Hansson:2018jt,Villois:2019kt,Erkintalo:2019wf,Lobanov:2020kt} or OPOs~\cite{Villois:2019gl,ParraRivas:2019gf}.

Finally, optical frequency combs are attracting a growing interest as a source of complex quantum states of light for high-dimensional quantum computation~\cite{Menicucci:2008ge,Kues:2019dh,Pfister:2020ck}.  
Second-order nonlinear optical systems are efficiently used for generation of quantum states of light: the classical correlations that establish in three-wave-mixing processes hold at the quantum level as well, leading, for instance, to generation of squeezed light or bipartite entanglement in an OPO.
Tripartite, or quadripartite multicolor entanglement has been predicted in second-order nonlinear devices\cite{Villar:2006ji}, in particular when multiple cascaded second-order nonlinear interactions occur, in traveling-wave or intracavity processes~\cite{Pfister:2004fs,Guo:2005cg,Pennarun:2007hl}.
Interestingly, a recent study based on the three-wave model of Eq.~(\ref{eq:d-FWM-a})-(\ref{eq:d-FWM-c}) predicts five-partite entanglement between one-octave-distant modes~\cite{He:2017gh}. 
This result suggests that quadratic combs could exhibit multipartite entanglement between frequency modes, which are essential for scalable measurement-based quantum computing~\cite{Raussendorf:2003ca}.  To fully explore these feature, a general and complete analysis of the quantum dynamics of quadratic comb is needed~\cite{Chembo:2016ki}.

\vspace{6pt} 



\authorcontributions{Designed and performed the experiments, I.R., S.M., M.P., P.M., and M.D.R.; theoretical modeling I.R, F.L., T.H., M.E., S.W., and M.D.R.; writing–original draft preparation, I.R. and M.D.R.; funding acquisition, M.E., P.D.N., S.W., and M.D.R.; all authors analyzed the data, discussed the results, read and edited the manuscript. All authors have read and agreed to the published version of the manuscript.}

\funding{
This research was funded by 
Ministero dell’Istruzione, dell’Università e della Ricerca (MIUR), PRIN 2015KEZNYM (NEMO); 
Ministero degli Affari Esteri e della Cooperazione Internazionale, project NOICE Joint Laboratory; 
European Union’s Horizon 2020 research and innovation programme (Qombs Project, FET Flagship on Quantum Technologies grant no. 820419);
the Rutherford Discovery Fellowships of the Royal Society of New Zealand and the Marsden Fund of the Royal Society of New Zealand.  
The work of S. W. is supported by the Ministry of Education and Science of the Russian Federation (Minobrnauka) (14.Y26.31.0017). 
T.H. acknowledges funding from the Swedish Research Council (Grant No. 2017-05309).
}

\acknowledgments{
}

\conflictsofinterest{The authors declare no conflict of interest.The funders had no role in the design of the study; in the collection, analyses, or interpretation of data; in the writing of the manuscript, or in the decision to publish the results.} 

\abbreviations{The following abbreviations are used in this manuscript:\\

\noindent 
\begin{tabular}{@{}ll}
cw & Continuous wave \\ 

DFG & Difference frequency generation \\

FF & Fundamental frequency \\

FFT & Fast Fourier transform \\

FSR & Free spectral range \\

FWM & Four-wave mixing \\

GVD & Group velocity dispersion\\

MI & Modulation instability\\

OFC & Optical frequency comb \\

OPO & Optical parametric oscillator\\

SH &Second harmonic\\

SHG & Second harmonic generation \\

TH & Third harmonic
\end{tabular}}

%

\reftitle{References}

\externalbibliography{yes}
\bibliography{Micromachines}

\begin{thebibliography}{-------}
\providecommand{\natexlab}[1]{#1}

\bibitem[Jones \em{et~al.}(2000)Jones, Diddams, Ranka, Stentz, Windeler, Hall,
  and Cundiff]{Jones:2000tn}
Jones, D.J.; Diddams, S.A.; Ranka, J.K.; Stentz, A.J.; Windeler, R.S.; Hall,
  J.L.; Cundiff, S.T.
\newblock {Carrier-envelope phase control of femtosecond mode-locked lasers and
  direct optical frequency synthesis}.
\newblock {\em Science} {\bf 2000}, {\em 288},~635--639.

\bibitem[Holzwarth \em{et~al.}(2000)Holzwarth, Udem, H\"ansch, Knight,
  Wadsworth, and Russell]{Holzwarth:2000aa}
Holzwarth, R.; Udem, T.; H\"ansch, T.W.; Knight, J.C.; Wadsworth, W.J.;
  Russell, P.S.J.
\newblock {Optical frequency synthesizer for precision spectroscopy}.
\newblock {\em Phys. Rev. Lett.} {\bf 2000}, {\em 85},~2264--2267.

\bibitem[Hall(2006)]{Hall:2006by}
Hall, J.L.
\newblock {Nobel Lecture: Defining and measuring optical frequencies}.
\newblock {\em Rev. Mod. Phys.} {\bf 2006}, {\em 78},~1279--1295.

\bibitem[Hansch(2006)]{Hansch:2006el}
Hansch, T.W.
\newblock {Nobel Lecture: Passion for precision}.
\newblock {\em Rev. Mod. Phys.} {\bf 2006}, {\em 78},~1297--1309.

\bibitem[Newbury(2011)]{Newbury:2011dh}
Newbury, N.R.
\newblock {Searching for applications with a fine-tooth comb}.
\newblock {\em Nat. Phot.} {\bf 2011}, {\em 5},~186--188.

\bibitem[Predehl \em{et~al.}(2012)Predehl, Grosche, Raupach, Droste, Terra,
  Alnis, Legero, Hansch, Udem, Holzwarth, and Schnatz]{Predehl:2012kr}
Predehl, K.; Grosche, G.; Raupach, S.M.F.; Droste, S.; Terra, O.; Alnis, J.;
  Legero, T.; Hansch, T.W.; Udem, T.; Holzwarth, R.; Schnatz, H.
\newblock {A 920-Kilometer Optical Fiber Link for Frequency Metrology at the
  19th Decimal Place}.
\newblock {\em Science} {\bf 2012}, {\em 336},~441--444.

\bibitem[Clivati \em{et~al.}(2016)Clivati, Cappellini, Livi, Poggiali,
  de~Cumis, Mancini, Pagano, Frittelli, Mura, Costanzo, Levi, Calonico,
  Fallani, Catani, and Inguscio]{Clivati:2016ij}
Clivati, C.; Cappellini, G.; Livi, L.F.; Poggiali, F.; de~Cumis, M.S.; Mancini,
  M.; Pagano, G.; Frittelli, M.; Mura, A.; Costanzo, G.A.; Levi, F.; Calonico,
  D.; Fallani, L.; Catani, J.; Inguscio, M.
\newblock {Measuring absolute frequencies beyond the GPS limit via long-haul
  optical frequency dissemination}.
\newblock {\em Opt. Express} {\bf 2016}, {\em 24},~11865--11875.

\bibitem[Insero \em{et~al.}(2017)Insero, Borri, Calonico, Pastor, Clivati,
  D'Ambrosio, De~Natale, Inguscio, Levi, and Santambrogio]{Insero:2017gg}
Insero, G.; Borri, S.; Calonico, D.; Pastor, P.C.; Clivati, C.; D'Ambrosio, D.;
  De~Natale, P.; Inguscio, M.; Levi, F.; Santambrogio, G.
\newblock Measuring molecular frequencies in the 1-10 $\mu$m range at 11-digits
  accuracy.
\newblock {\em Sci. Rep.} {\bf 2017}, {\em 7},~12780.

\bibitem[Di~Sarno \em{et~al.}(2019)Di~Sarno, Aiello, De~Rosa, Ricciardi, Mosca,
  Notariale, De~Natale, Santamaria, and Maddaloni]{DiSarno:2019ep}
Di~Sarno, V.; Aiello, R.; De~Rosa, M.; Ricciardi, I.; Mosca, S.; Notariale, G.;
  De~Natale, P.; Santamaria, L.; Maddaloni, P.
\newblock {Lamb-dip spectroscopy of buffer-gas-cooled molecules}.
\newblock {\em Optica} {\bf 2019}, {\em 6},~436--441.

\bibitem[Steinmetz \em{et~al.}(2008)Steinmetz, Wilken, Araujo-Hauck, Holzwarth,
  Haensch, Pasquini, Manescau, D'Odorico, Murphy, Kentischer, Schmidt, and
  Udem]{Steinmetz:2008he}
Steinmetz, T.; Wilken, T.; Araujo-Hauck, C.; Holzwarth, R.; Haensch, T.W.;
  Pasquini, L.; Manescau, A.; D'Odorico, S.; Murphy, M.T.; Kentischer, T.;
  Schmidt, W.; Udem, T.
\newblock {Laser frequency combs for astronomical observations}.
\newblock {\em Science} {\bf 2008}, {\em 321},~1335--1337.

\bibitem[McCracken \em{et~al.}(2017)McCracken, Charsley, and
  Reid]{McCracken:2017dy}
McCracken, R.A.; Charsley, J.M.; Reid, D.T.
\newblock {A decade of astrocombs: recent advances in frequency combs for
  astronomy}.
\newblock {\em Opt. Express} {\bf 2017}, {\em 25},~15058--15078.

\bibitem[Obrzud \em{et~al.}(2018)Obrzud, Rainer, Harutyunyan, Anderson, Liu,
  Geiselmann, Chazelas, Kundermann, Lecomte, Cecconi, Ghedina, Molinari, Pepe,
  Wildi, Bouchy, Kippenberg, and Herr]{Obrzud:2018cb}
Obrzud, E.; Rainer, M.; Harutyunyan, A.; Anderson, M.H.; Liu, J.; Geiselmann,
  M.; Chazelas, B.; Kundermann, S.; Lecomte, S.; Cecconi, M.; Ghedina, A.;
  Molinari, E.; Pepe, F.; Wildi, F.; Bouchy, F.; Kippenberg, T.J.; Herr, T.
\newblock {A microphotonic astrocomb}.
\newblock {\em Nature Photon.} {\bf 2018}, {\em 13},~31--35.

\bibitem[Adler \em{et~al.}(2010)Adler, Thorpe, Cossel, and Ye]{Adler:2010da}
Adler, F.; Thorpe, M.J.; Cossel, K.C.; Ye, J.
\newblock {Cavity-Enhanced Direct Frequency Comb Spectroscopy: Technology and
  Applications}.
\newblock {\em Annu. Rev. Anal. Chem.} {\bf 2010}, {\em 3},~175--205.

\bibitem[Keilmann \em{et~al.}(2004)Keilmann, Gohle, and
  Holzwarth]{Keilmann:2004gd}
Keilmann, F.; Gohle, C.; Holzwarth, R.
\newblock {Time-domain mid-infrared frequency-comb spectrometer.}
\newblock {\em Opt. Lett.} {\bf 2004}, {\em 29},~1542--1544.

\bibitem[Picqu{\'e} and Hansch(2019)]{Picque:2019jz}
Picqu{\'e}, N.; Hansch, T.W.
\newblock {Frequency comb spectroscopy}.
\newblock {\em Nature Photon.} {\bf 2019}, {\em 13},~146--157.

\bibitem[Schliesser \em{et~al.}(2012)Schliesser, Picqu{\'e}, and
  H{\"a}nsch]{Schliesser:2012dn}
Schliesser, A.; Picqu{\'e}, N.; H{\"a}nsch, T.W.
\newblock {Mid-infrared frequency combs}.
\newblock {\em Nature Photon.} {\bf 2012}, {\em 6},~440--449.

\bibitem[Rieker \em{et~al.}(2014)Rieker, Giorgetta, Swann, Kofler, Zolot,
  Sinclair, Baumann, Cromer, Petron, Sweeney, Tans, Coddington, and
  Newbury]{Rieker:2014fj}
Rieker, G.B.; Giorgetta, F.R.; Swann, W.C.; Kofler, J.; Zolot, A.M.; Sinclair,
  L.C.; Baumann, E.; Cromer, C.; Petron, G.; Sweeney, C.; Tans, P.P.;
  Coddington, I.; Newbury, N.R.
\newblock {Frequency-comb-based remote sensing of greenhouse gases over
  kilometer air paths}.
\newblock {\em Optica} {\bf 2014}, {\em 1},~290--298.

\bibitem[Yu \em{et~al.}(2018)Yu, Okawachi, Griffith, Picqu{\'e}, Lipson, and
  Gaeta]{Yu:2018ku}
Yu, M.; Okawachi, Y.; Griffith, A.G.; Picqu{\'e}, N.; Lipson, M.; Gaeta, A.L.
\newblock {Silicon-chip-based mid-infrared dual-comb spectroscopy}.
\newblock {\em Nature Commun.} {\bf 2018}, {\em 9},~1869.

\bibitem[Pfeifle \em{et~al.}(2014)Pfeifle, Brasch, Lauermann, Yu, Wegner, Herr,
  Hartinger, Schindler, Li, Hillerkuss, Schmogrow, Weimann, Holzwarth, Freude,
  Leuthold, Kippenberg, and Koos]{Pfeifle:2014cm}
Pfeifle, J.; Brasch, V.; Lauermann, M.; Yu, Y.; Wegner, D.; Herr, T.;
  Hartinger, K.; Schindler, P.; Li, J.; Hillerkuss, D.; Schmogrow, R.; Weimann,
  C.; Holzwarth, R.; Freude, W.; Leuthold, J.; Kippenberg, T.J.; Koos, C.
\newblock {Coherent terabit communications with microresonator Kerr frequency
  combs}.
\newblock {\em Nat. Phot.} {\bf 2014}, {\em 8},~375--380.

\bibitem[Kemal \em{et~al.}(2016)Kemal, Pfeifle, Marin-Palomo, Pascual, Wolf,
  Smyth, Freude, and Koos]{Kemal:2016ho}
Kemal, J.N.; Pfeifle, J.; Marin-Palomo, P.; Pascual, M.D.G.; Wolf, S.; Smyth,
  F.; Freude, W.; Koos, C.
\newblock {Multi-wavelength coherent transmission using an optical frequency
  comb as a local oscillator}.
\newblock {\em Opt. Express} {\bf 2016}, {\em 24},~25432--25445.

\bibitem[Marin-Palomo \em{et~al.}(2017)Marin-Palomo, Kemal, Karpov, Kordts,
  Pfeifle, Pfeiffer, Trocha, Wolf, Brasch, Anderson, Rosenberger, Vijayan,
  Freude, Kippenberg, and Koos]{MarinPalomo:2017bv}
Marin-Palomo, P.; Kemal, J.N.; Karpov, M.; Kordts, A.; Pfeifle, J.; Pfeiffer,
  M.H.P.; Trocha, P.; Wolf, S.; Brasch, V.; Anderson, M.H.; Rosenberger, R.;
  Vijayan, K.; Freude, W.; Kippenberg, T.J.; Koos, C.
\newblock {Microresonator-based solitons for massively parallel coherent
  optical communications}.
\newblock {\em Nature} {\bf 2017}, {\em 546},~274--279.

\bibitem[Roslund \em{et~al.}(2014)Roslund, de~Ara{\'u}jo, Jiang, Fabre, and
  Treps]{Roslund:2014cb}
Roslund, J.; de~Ara{\'u}jo, R.M.; Jiang, S.; Fabre, C.; Treps, N.
\newblock {Wavelength-multiplexed quantum networks with ultrafast frequency
  combs}.
\newblock {\em Nature Photon.} {\bf 2014}, {\em 8},~109--112.

\bibitem[Dutt \em{et~al.}(2015)Dutt, Luke, Manipatruni, Gaeta, Nussenzveig, and
  Lipson]{Dutt:2015fx}
Dutt, A.; Luke, K.; Manipatruni, S.; Gaeta, A.L.; Nussenzveig, P.; Lipson, M.
\newblock {On-Chip Optical Squeezing}.
\newblock {\em Phys. Rev. Applied} {\bf 2015}, {\em 3},~044005.

\bibitem[Reimer \em{et~al.}(2016)Reimer, Kues, Roztocki, Wetzel, Grazioso,
  Little, Chu, Johnston, Bromberg, Caspani, Moss, and
  Morandotti]{Reimer:2016jk}
Reimer, C.; Kues, M.; Roztocki, P.; Wetzel, B.; Grazioso, F.; Little, B.E.;
  Chu, S.T.; Johnston, T.; Bromberg, Y.; Caspani, L.; Moss, D.J.; Morandotti,
  R.
\newblock {Generation of multiphoton entangled quantum states by means of
  integrated frequency combs}.
\newblock {\em Science} {\bf 2016}, {\em 351},~1176--1180.

\bibitem[Imany \em{et~al.}(2018)Imany, Jaramillo-Villegas, Odele, Kyunghun,
  Leaird, Lukens, Lougovski, Minghao, and Weiner]{Imany:2018jk}
Imany, P.; Jaramillo-Villegas, J.A.; Odele, O.D.; Kyunghun, H.A.N.; Leaird,
  D.E.; Lukens, J.M.; Lougovski, P.; Minghao, Q.I.; Weiner, A.M.
\newblock {50-GHz-spaced comb of high-dimensional frequency-bin entangled
  photons from an on-chip silicon nitride microresonator}.
\newblock {\em Opt. Express} {\bf 2018}, {\em 26},~1825--1840.

\bibitem[Kues \em{et~al.}(2019)Kues, Reimer, Lukens, Munro, Weiner, Moss, and
  Morandotti]{Kues:2019dh}
Kues, M.; Reimer, C.; Lukens, J.M.; Munro, W.J.; Weiner, A.M.; Moss, D.J.;
  Morandotti, R.
\newblock {Quantum optical microcombs}.
\newblock {\em Nature Photon.} {\bf 2019}, {\em 13},~170--179.

\bibitem[Del'Haye \em{et~al.}(2007)Del'Haye, Schliesser, Arcizet, Wilken,
  Holzwarth, and Kippenberg]{DelHaye:2007gi}
Del'Haye, P.; Schliesser, A.; Arcizet, O.; Wilken, T.; Holzwarth, R.;
  Kippenberg, T.J.
\newblock {Optical frequency comb generation from a monolithic microresonator}.
\newblock {\em Nature} {\bf 2007}, {\em 450},~1214--1217.

\bibitem[Kippenberg \em{et~al.}(2011)Kippenberg, Holzwarth, and
  Diddams]{Kippenberg:2011fc}
Kippenberg, T.J.; Holzwarth, R.; Diddams, S.A.
\newblock {Microresonator-based optical frequency combs}.
\newblock {\em Science} {\bf 2011}, {\em 332},~555--559.

\bibitem[Pasquazi \em{et~al.}(2018)Pasquazi, Peccianti, Razzari, Moss, Coen,
  Erkintalo, Chembo, Hansson, Wabnitz, Del'Haye, Xue, Weiner, and
  Morandotti]{Pasquazi:2018de}
Pasquazi, A.; Peccianti, M.; Razzari, L.; Moss, D.J.; Coen, S.; Erkintalo, M.;
  Chembo, Y.K.; Hansson, T.; Wabnitz, S.; Del'Haye, P.; Xue, X.; Weiner, A.M.;
  Morandotti, R.
\newblock {Micro-combs: A novel generation of optical sources}.
\newblock {\em Phys. Rep.} {\bf 2018}, {\em 729},~1--81.

\bibitem[Gaeta \em{et~al.}(2019)Gaeta, Lipson, and Kippenberg]{Gaeta:2019hc}
Gaeta, A.L.; Lipson, M.; Kippenberg, T.J.
\newblock {Photonic-chip-based frequency combs}.
\newblock {\em Nature Photon.} {\bf 2019}, {\em 13},~158--169.

\bibitem[Maddaloni \em{et~al.}(2006)Maddaloni, Malara, Gagliardi, and
  De~Natale]{Maddaloni:2006ka}
Maddaloni, P.; Malara, P.; Gagliardi, G.; De~Natale, P.
\newblock {Mid-infrared fibre-based optical comb}.
\newblock {\em New J. Phys.} {\bf 2006}, {\em 8},~262--262.

\bibitem[Sun \em{et~al.}(2007)Sun, Gale, and Reid]{Sun:2007wu}
Sun, J.H.; Gale, B.J.S.; Reid, D.T.
\newblock {Composite frequency comb spanning 0.4-2.4 $\mu$m from a
  phase-controlled femtosecond Ti:sapphire laser and synchronously pumped
  optical parametric oscillator.}
\newblock {\em Opt. Lett.} {\bf 2007}, {\em 32},~1414--1416.

\bibitem[Wong \em{et~al.}(2008)Wong, Plettner, Vodopyanov, Urbanek, Digonnet,
  and Byer]{Wong:2008fz}
Wong, S.T.; Plettner, T.; Vodopyanov, K.L.; Urbanek, K.; Digonnet, M.; Byer,
  R.L.
\newblock {Self-phase-locked degenerate femtosecond optical parametric
  oscillator.}
\newblock {\em Opt. Lett.} {\bf 2008}, {\em 33},~1896--1898.

\bibitem[Gambetta \em{et~al.}(2008)Gambetta, Ramponi, and
  Marangoni]{Gambetta:2008fa}
Gambetta, A.; Ramponi, R.; Marangoni, M.
\newblock {Mid-infrared optical combs from a compact amplified Er-doped fiber
  oscillator.}
\newblock {\em Opt. Lett.} {\bf 2008}, {\em 33},~2671--2673.

\bibitem[Adler \em{et~al.}(2009)Adler, Cossel, Thorpe, Hartl, Fermann, and
  Ye]{Adler:2009ka}
Adler, F.; Cossel, K.C.; Thorpe, M.J.; Hartl, I.; Fermann, M.E.; Ye, J.
\newblock {Phase-stabilized, 15 W frequency comb at 2.8--4.8~$\mu$m}.
\newblock {\em Opt. Lett.} {\bf 2009}, {\em 34},~1330--1332.

\bibitem[Galli \em{et~al.}(2013)Galli, Cappelli, Cancio, Giusfredi, Mazzotti,
  Bartalini, and De~Natale]{Galli:2013cg}
Galli, I.; Cappelli, F.; Cancio, P.; Giusfredi, G.; Mazzotti, D.; Bartalini,
  S.; De~Natale, P.
\newblock {High-coherence mid-infrared frequency comb}.
\newblock {\em Opt. Express} {\bf 2013}, {\em 21},~28877--28885.

\bibitem[Diddams \em{et~al.}(1999)Diddams, Ma, Ye, and Hall]{Diddams:1999ch}
Diddams, S.A.; Ma, L.S.S.; Ye, J.; Hall, J.L.
\newblock {Broadband optical frequency comb generation with a phase-modulated
  parametric oscillator}.
\newblock {\em Opt. Lett.} {\bf 1999}, {\em 24},~1747--1749.

\bibitem[{Kourogi} \em{et~al.}(1993){Kourogi}, {Nakagawa}, and
  {Ohtsu}]{Kourogi:1993gy}
{Kourogi}, M.; {Nakagawa}, K.; {Ohtsu}, M.
\newblock Wide-span optical frequency comb generator for accurate optical
  frequency difference measurement.
\newblock {\em IEEE J. Quantum Electron.} {\bf 1993}, {\em 29},~2693--2701.

\bibitem[Ulvila \em{et~al.}(2013)Ulvila, Phillips, Halonen, and
  Vainio]{Ulvila:2013jv}
Ulvila, V.; Phillips, C.R.; Halonen, L.L.; Vainio, M.
\newblock {Frequency comb generation by a continuous-wave-pumped optical
  parametric oscillator based on cascading quadratic nonlinearities}.
\newblock {\em Opt. Lett.} {\bf 2013}, {\em 38},~4281--4284.

\bibitem[Ulvila \em{et~al.}(2014)Ulvila, Phillips, Halonen, and
  Vainio]{Ulvila:2014bx}
Ulvila, V.; Phillips, C.R.; Halonen, L.L.; Vainio, M.
\newblock {High-power mid-infrared frequency comb from a continuous-wave-pumped
  bulk optical parametric oscillator}.
\newblock {\em Opt. Express} {\bf 2014}, {\em 22},~10535--10543.

\bibitem[Ostrovskii(1967)]{Ostrovskii:1967}
Ostrovskii, L.A.
\newblock {Self-action of Light in Crystals}.
\newblock {\em JETP Lett.} {\bf 1967}, {\em 5},~272--275.

\bibitem[Desalvo \em{et~al.}(1992)Desalvo, Hagan, Sheik-Bahae, Stegeman,
  Van~Stryland, and Vanherzeele]{Desalvo:1992cs}
Desalvo, R.; Hagan, D.J.; Sheik-Bahae, M.; Stegeman, G.; Van~Stryland, E.W.;
  Vanherzeele, H.
\newblock {Self-focusing and self-defocusing by cascaded second-order effects
  in KTP}.
\newblock {\em Opt. Lett.} {\bf 1992}, {\em 17},~28--30.

\bibitem[Stegeman(1999)]{Stegeman:1999fe}
Stegeman, G.I.
\newblock {$\chi^{(2)}$ cascading: nonlinear phase shifts}.
\newblock {\em Quantum Semiclass. Opt.} {\bf 1999}, {\em 9},~139--153.

\bibitem[Schiller and Byer(1993)]{Schiller:1993tz}
Schiller, S.; Byer, R.L.
\newblock {Quadruply resonant optical parametric oscillation in a monolithic
  total-internal-reflection resonator}.
\newblock {\em J. Opt. Soc. Am. B} {\bf 1993}, {\em 10},~1696--1707.

\bibitem[Schiller \em{et~al.}(1996)Schiller, Breitenbach, Paschotta, and
  Mlynek]{Schiller:1996gx}
Schiller, S.; Breitenbach, G.; Paschotta, R.R.; Mlynek, J.
\newblock {Subharmonic-pumped continuous-wave parametric oscillator}.
\newblock {\em Appl. Phys. Lett.} {\bf 1996}, {\em 68},~3374--3376.

\bibitem[Schneider and Schiller(1997)]{Schneider:1997ks}
Schneider, K.; Schiller, S.
\newblock {Multiple conversion and optical limiting in a subharmonic-pumped
  parametric oscillator}.
\newblock {\em Opt. Lett.} {\bf 1997}, {\em 22},~363--365.

\bibitem[White \em{et~al.}(1997)White, Lam, Taubman, Marte, Schiller,
  McClelland, and Bachor]{White:1997ta}
White, A.G.; Lam, P.K.; Taubman, M.S.; Marte, M.A.M.; Schiller, S.; McClelland,
  D.E.; Bachor, H.A.
\newblock {Classical and quantum signatures of competing $\chi^{(2)}$
  nonlinearities}.
\newblock {\em Phys. Rev. A} {\bf 1997}, {\em 55},~4511--4515.

\bibitem[Ricciardi \em{et~al.}(2015)Ricciardi, Mosca, Parisi, Maddaloni,
  Santamaria, De~Natale, and De~Rosa]{Ricciardi:2015bw}
Ricciardi, I.; Mosca, S.; Parisi, M.; Maddaloni, P.; Santamaria, L.; De~Natale,
  P.; De~Rosa, M.
\newblock {Frequency comb generation in quadratic nonlinear media}.
\newblock {\em Phys. Rev. A} {\bf 2015}, {\em 91},~063839.

\bibitem[Mosca \em{et~al.}(2016)Mosca, Ricciardi, Parisi, Maddaloni,
  Santamaria, De~Natale, and De~Rosa]{Mosca:2015wh}
Mosca, S.; Ricciardi, I.; Parisi, M.; Maddaloni, P.; Santamaria, L.; De~Natale,
  P.; De~Rosa, M.
\newblock {Direct generation of optical frequency combs in $\chi^{(2)}$
  nonlinear cavities}.
\newblock {\em Nanophotonics} {\bf 2016}, {\em 5},~316--331.

\bibitem[Leo \em{et~al.}(2016{\natexlab{a}})Leo, Hansson, Ricciardi, De~Rosa,
  Coen, Wabnitz, and Erkintalo]{Leo:2016kj}
Leo, F.; Hansson, T.; Ricciardi, I.; De~Rosa, M.; Coen, S.; Wabnitz, S.;
  Erkintalo, M.
\newblock {Walk-off-induced modulation instability, temporal pattern formation,
  and frequency comb generation in cavity-enhanced second-harmonic generation}.
\newblock {\em Phys. Rev. Lett.} {\bf 2016}, {\em 116},~{033901}.

\bibitem[Leo \em{et~al.}(2016{\natexlab{b}})Leo, Hansson, Ricciardi, De~Rosa,
  Coen, Wabnitz, and Erkintalo]{Leo:2016df}
Leo, F.; Hansson, T.; Ricciardi, I.; De~Rosa, M.; Coen, S.; Wabnitz, S.;
  Erkintalo, M.
\newblock {Frequency-comb formation in doubly resonant second-harmonic
  generation}.
\newblock {\em Phys. Rev. A} {\bf 2016}, {\em 93},~043831.

\bibitem[Hansson \em{et~al.}(2017)Hansson, Leo, Erkintalo, Coen, Ricciardi,
  De~Rosa, and Wabnitz]{Hansson:2017cs}
Hansson, T.; Leo, F.; Erkintalo, M.; Coen, S.; Ricciardi, I.; De~Rosa, M.;
  Wabnitz, S.
\newblock {Singly resonant second-harmonic-generation frequency combs}.
\newblock {\em Phys. Rev. A} {\bf 2017}, {\em 95},~013805.

\bibitem[Zakharov and Ostrovsky(2009)]{Zakharov:2009du}
Zakharov, V.E.; Ostrovsky, L.A.
\newblock {Modulation instability: The beginning}.
\newblock {\em Physica D: Nonlinear Phenomena} {\bf 2009}, {\em 238},~540--548.

\bibitem[Mosca \em{et~al.}(2018)Mosca, Parisi, Ricciardi, Leo, Hansson,
  Erkintalo, Maddaloni, De~Natale, Wabnitz, and De~Rosa]{Mosca:2018jk}
Mosca, S.; Parisi, M.; Ricciardi, I.; Leo, F.; Hansson, T.; Erkintalo, M.;
  Maddaloni, P.; De~Natale, P.; Wabnitz, S.; De~Rosa, M.
\newblock {Modulation Instability Induced Frequency Comb Generation in a
  Continuously Pumped Optical Parametric Oscillator}.
\newblock {\em Phys. Rev. Lett.} {\bf 2018}, {\em 121},~093903.

\bibitem[Hansson \em{et~al.}(2016)Hansson, Leo, Erkintalo, Anthony, Coen,
  Ricciardi, De~Rosa, and Wabnitz]{Hansson:2016kz}
Hansson, T.; Leo, F.; Erkintalo, M.; Anthony, J.; Coen, S.; Ricciardi, I.;
  De~Rosa, M.; Wabnitz, S.
\newblock {Single envelope equation modeling of multi-octave comb arrays in
  microresonators with quadratic and cubic nonlinearities}.
\newblock {\em J. Opt. Soc. Am. B} {\bf 2016}, {\em 33},~1207--1215.

\bibitem[Drever \em{et~al.}(1983)Drever, Hall, Kowalski, Hough, Ford, Munley,
  and Ward]{Drever:1983gx}
Drever, R.W.P.; Hall, J.L.; Kowalski, F.V.; Hough, J.; Ford, G.M.; Munley,
  A.J.; Ward, H.
\newblock {Laser phase and frequency stabilization using an optical resonator}.
\newblock {\em Appl. Phys. B} {\bf 1983}, {\em 31},~97--105.

\bibitem[Ricciardi \em{et~al.}(2010)Ricciardi, De~Rosa, Rocco, Ferraro, and
  De~Natale]{Ricciardi:2010kd}
Ricciardi, I.; De~Rosa, M.; Rocco, A.; Ferraro, P.; De~Natale, P.
\newblock {Cavity-enhanced generation of 6 W cw second-harmonic power at 532 nm
  in periodically-poled MgO:LiTaO$_3$}.
\newblock {\em Opt. Express} {\bf 2010}, {\em 18},~10985--10994.

\bibitem[De~Rosa \em{et~al.}(2002)De~Rosa, Conti, Cerdonio, Pinard, and
  Marin]{DeRosa:2002cg}
De~Rosa, M.; Conti, L.; Cerdonio, M.; Pinard, M.; Marin, F.
\newblock {Experimental Measurement of the Dynamic Photothermal Effect in
  Fabry-Perot Cavities for Gravitational Wave Detectors}.
\newblock {\em Phys. Rev. Lett.} {\bf 2002}, {\em 89},~237402.

\bibitem[Carmon \em{et~al.}(2004)Carmon, Yang, and Vahala]{Carmon:2004us}
Carmon, T.; Yang, L.; Vahala, K.J.
\newblock {Dynamical thermal behavior and thermal self-stability of
  microcavities}.
\newblock {\em Opt. Express} {\bf 2004}, {\em 12},~4742--4750.

\bibitem[Chembo and Yu(2010)]{Chembo:2010cb}
Chembo, Y.K.; Yu, N.
\newblock {Modal expansion approach to optical-frequency-comb generation with
  monolithic whispering-gallery-mode resonators}.
\newblock {\em Phys. Rev. A} {\bf 2010}, {\em 82},~033801.

\bibitem[Chembo \em{et~al.}(2010)Chembo, Strekalov, and Yu]{Chembo:2010ii}
Chembo, Y.K.; Strekalov, D.V.; Yu, N.
\newblock {Spectrum and dynamics of optical frequency combs generated with
  monolithic whispering gallery mode resonators}.
\newblock {\em Phys. Rev. Lett.} {\bf 2010}, {\em 104},~103902.

\bibitem[Haelterman \em{et~al.}(1992)Haelterman, Trillo, and
  Wabnitz]{Haelterman:1992cd}
Haelterman, M.; Trillo, S.; Wabnitz, S.
\newblock {Dissipative modulation instability in a nonlinear dispersive ring
  cavity}.
\newblock {\em Opt. Commun.} {\bf 1992}, {\em 91},~401--407.

\bibitem[Leo \em{et~al.}(2010)Leo, Coen, Kockaert, Gorza, Emplit, and
  Haelterman]{Leo:2010if}
Leo, F.; Coen, S.; Kockaert, P.; Gorza, S.P.; Emplit, P.; Haelterman, M.
\newblock {Temporal cavity solitons in one-dimensional Kerr media as bits in an
  all-optical buffer}.
\newblock {\em Nature Photon.} {\bf 2010}, {\em 4},~471--476.

\bibitem[Coen \em{et~al.}(2013)Coen, Randle, Sylvestre, and
  Erkintalo]{Coen:2013hw}
Coen, S.; Randle, H.G.; Sylvestre, T.; Erkintalo, M.
\newblock {Modeling of octave-spanning Kerr frequency combs using a generalized
  mean-field Lugiato--Lefever model}.
\newblock {\em Opt. Lett.} {\bf 2013}, {\em 38},~37--39.

\bibitem[Coen and Erkintalo(2013)]{Coen:2013hd}
Coen, S.; Erkintalo, M.
\newblock {Universal scaling laws of Kerr frequency combs}.
\newblock {\em Opt. Lett.} {\bf 2013}, {\em 38},~1790--1792.

\bibitem[Hansson \em{et~al.}(2014)Hansson, Modotto, and
  Wabnitz]{Hansson:2014ie}
Hansson, T.; Modotto, D.; Wabnitz, S.
\newblock {On the numerical simulation of Kerr frequency combs using coupled
  mode equations}.
\newblock {\em Opt. Commun.} {\bf 2014}, {\em 312},~134--136.

\bibitem[Agrawal(2001)]{Agrawal:2001book}
Agrawal, G.P.
\newblock {\em Nonlinear Fiber Optics}, 3rd ed.; Academic Press: San Diego,
  2001.

\bibitem[Weideman \em{et~al.}(1986)Weideman, Herbst, and
  {1986}]{Weideman:1986hc}
Weideman, J.; Herbst, B.; {1986}.
\newblock {Split-step methods for the solution of the nonlinear Schr{\"o}dinger
  equation}.
\newblock {\em SIAM} {\bf 1986}, {\em 23},~485--507.

\bibitem[Godey \em{et~al.}(2014)Godey, Balakireva, Coillet, and
  Chembo]{Godey:2014ks}
Godey, C.; Balakireva, I.V.; Coillet, A.; Chembo, Y.K.
\newblock {Stability analysis of the spatiotemporal Lugiato-Lefever model for
  Kerr optical frequency combs in the anomalous and normal dispersion regimes}.
\newblock {\em Phys. Rev. A} {\bf 2014}.

\bibitem[Khurgin \em{et~al.}(2014)Khurgin, Dikmelik, Hugi, and
  Faist]{Khurgin:2014hy}
Khurgin, J.B.; Dikmelik, Y.; Hugi, A.; Faist, J.
\newblock {Coherent frequency combs produced by self frequency modulation in
  quantum cascade lasers}.
\newblock {\em Appl. Phys. Lett.} {\bf 2014}, {\em 104},~081118.

\bibitem[Cappelli \em{et~al.}(2019)Cappelli, Consolino, Campo, Galli, Mazzotti,
  Campa, de~Cumis, Pastor, Eramo, R{\"o}sch, Beck, Scalari, Faist, De~Natale,
  and Bartalini]{Cappelli:2019bn}
Cappelli, F.; Consolino, L.; Campo, G.; Galli, I.; Mazzotti, D.; Campa, A.;
  de~Cumis, M.S.; Pastor, P.C.; Eramo, R.; R{\"o}sch, M.; Beck, M.; Scalari,
  G.; Faist, J.; De~Natale, P.; Bartalini, S.
\newblock {Retrieval of phase relation and emission profile of quantum cascade
  laser frequency combs}.
\newblock {\em Nature Photon.} {\bf 2019}, {\em 288},~1.

\bibitem[Breunig(2016)]{Breunig:2016ku}
Breunig, I.
\newblock {Three-wave mixing in whispering gallery resonators}.
\newblock {\em Laser {\&} Photon. Rev.} {\bf 2016}, {\em 10},~569--587.

\bibitem[Strekalov \em{et~al.}(2016)Strekalov, Marquardt, Matsko, Schwefel, and
  Leuchs]{Strekalov:2016iw}
Strekalov, D.V.; Marquardt, C.; Matsko, A.B.; Schwefel, H.G.L.; Leuchs, G.
\newblock {Nonlinear and quantum optics with whispering gallery resonators}.
\newblock {\em J. Opt.} {\bf 2016}, {\em 18},~123002.

\bibitem[Ikuta \em{et~al.}(2018)Ikuta, Asano, Tani, Yamamoto, and
  Imoto]{Ikuta:2018iw}
Ikuta, R.; Asano, M.; Tani, R.; Yamamoto, T.; Imoto, N.
\newblock {Frequency comb generation in a quadratic nonlinear waveguide
  resonator}.
\newblock {\em Opt. Express} {\bf 2018}, {\em 26},~15551--15558.

\bibitem[Stefszky \em{et~al.}(2018)Stefszky, Ulvila, Abdallah, Silberhorn, and
  Vainio]{Stefszky:2018bd}
Stefszky, M.; Ulvila, V.; Abdallah, Z.; Silberhorn, C.; Vainio, M.
\newblock {Towards optical-frequency-comb generation in continuous-wave-pumped
  titanium-indiffused lithium-niobate waveguide resonators}.
\newblock {\em Phys. Rev. A} {\bf 2018}, {\em 98},~053850.

\bibitem[Hendry \em{et~al.}(2019)Hendry, Trainor, Xu, Coen, Murdoch, Schwefel,
  and Erkintalo]{Hendry:2019tb}
Hendry, I.; Trainor, L.S.; Xu, Y.; Coen, S.; Murdoch, S.G.; Schwefel, H.G.L.;
  Erkintalo, M.
\newblock {Experimental observation of internally-pumped parametric oscillation
  and quadratic comb generation in a $\chi^{(2)}$ whispering-gallery-mode
  microresonator}.
\newblock {\em arXiv:1912.02804} {\bf 2019},
  \href{http://xxx.lanl.gov/abs/1912.02804}{{\normalfont [1912.02804]}}.

\bibitem[Szabados \em{et~al.}(2019)Szabados, Puzyrev, Minet, Reis, Buse,
  Villois, Skryabin, and Breunig]{Szabados:2019tf}
Szabados, J.; Puzyrev, D.N.; Minet, Y.; Reis, L.; Buse, K.; Villois, A.;
  Skryabin, D.V.; Breunig, I.
\newblock {Frequency comb generation via cascaded second-order nonlinearities
  in microresonators}.
\newblock {\em arXiv:1912.00945} {\bf 2019},
  \href{http://xxx.lanl.gov/abs/1912.00945}{{\normalfont [1912.00945]}}.

\bibitem[Levy \em{et~al.}(2011)Levy, Foster, Gaeta, and Lipson]{Levy:2011wp}
Levy, J.S.; Foster, M.A.; Gaeta, A.L.; Lipson, M.
\newblock {Harmonic generation in silicon nitride ring resonators.}
\newblock {\em Opt. Express} {\bf 2011}, {\em 19},~11415--11421.

\bibitem[Jung \em{et~al.}(2014)Jung, Stoll, Guo, Fischer, and
  Tang]{Jung:2014jt}
Jung, H.; Stoll, R.; Guo, X.; Fischer, D.; Tang, H.X.
\newblock {Green, red, and IR frequency comb line generation from single IR
  pump in AlN microring resonator}.
\newblock {\em Optica} {\bf 2014}, {\em 1},~396--399.

\bibitem[Pu \em{et~al.}(2016)Pu, Ottaviano, Semenova, and Yvind]{Pu:2016en}
Pu, M.; Ottaviano, L.; Semenova, E.; Yvind, K.
\newblock {Efficient frequency comb generation in AlGaAs-on-insulator}.
\newblock {\em Optica} {\bf 2016}, {\em 3},~823--826.

\bibitem[Wang \em{et~al.}(2019)Wang, Zhang, Yu, Zhu, Hu, and Lon{\v
  c}ar]{Wang:2019hn}
Wang, C.; Zhang, M.; Yu, M.; Zhu, R.; Hu, H.; Lon{\v c}ar, M.
\newblock {Monolithic lithium niobate photonic circuits for Kerr frequency comb
  generation and modulation}.
\newblock {\em Nature Commun.} {\bf 2019}, {\em 10},~978.

\bibitem[Zhang \em{et~al.}(2019)Zhang, Buscaino, Wang, Shams-Ansari, Reimer,
  Zhu, Kahn, and Lon{\v c}ar]{Zhang:2019fe}
Zhang, M.; Buscaino, B.; Wang, C.; Shams-Ansari, A.; Reimer, C.; Zhu, R.; Kahn,
  J.M.; Lon{\v c}ar, M.
\newblock {Broadband electro-optic frequency comb generation in a lithium
  niobate microring resonator}.
\newblock {\em Nature} {\bf 2019}, {\em 568},~373--377.

\bibitem[Xue \em{et~al.}(2016)Xue, Leo, Xuan, Jaramillo-Villegas, Wang, Leaird,
  Erkintalo, Qi, and Weiner]{Xue:2016ja}
Xue, X.; Leo, F.; Xuan, Y.; Jaramillo-Villegas, J.A.; Wang, P.H.; Leaird, D.E.;
  Erkintalo, M.; Qi, M.; Weiner, A.M.
\newblock {Second-harmonic-assisted four-wave mixing in chip-based
  microresonator frequency comb generation}.
\newblock {\em Light Sci. Appl.} {\bf 2016}, {\em 6},~e16253.

\bibitem[F{\"u}rst \em{et~al.}(2010{\natexlab{a}})F{\"u}rst, Strekalov, Elser,
  Aiello, Andersen, Marquardt, and Leuchs]{Furst:2010bt}
F{\"u}rst, J.U.; Strekalov, D.V.; Elser, D.; Aiello, A.; Andersen, U.L.;
  Marquardt, C.; Leuchs, G.
\newblock {Low-Threshold Optical Parametric Oscillations in a Whispering
  Gallery Mode Resonator}.
\newblock {\em Phys. Rev. Lett.} {\bf 2010}, {\em 105},~263904.

\bibitem[F{\"u}rst \em{et~al.}(2010{\natexlab{b}})F{\"u}rst, Strekalov, Elser,
  Lassen, Andersen, Marquardt, and Leuchs]{Furst:2010co}
F{\"u}rst, J.U.; Strekalov, D.V.; Elser, D.; Lassen, M.; Andersen, U.L.;
  Marquardt, C.; Leuchs, G.
\newblock {Naturally Phase-Matched Second-Harmonic Generation in a
  Whispering-Gallery-Mode Resonator}.
\newblock {\em Phys. Rev. Lett.} {\bf 2010}, {\em 104},~153901.

\bibitem[Lin \em{et~al.}(2013)Lin, F{\"u}rst, Strekalov, and Yu]{Lin:2013fx}
Lin, G.; F{\"u}rst, J.U.; Strekalov, D.V.; Yu, N.
\newblock {Wide-range cyclic phase matching and second harmonic generation in
  whispering gallery resonators}.
\newblock {\em Appl. Phys. Lett.} {\bf 2013}, {\em 103},~181107.

\bibitem[Meisenheimer \em{et~al.}(2015)Meisenheimer, F{\"u}rst, Werner,
  Beckmann, Buse, and Breunig]{Meisenheimer:2015dn}
Meisenheimer, S.K.; F{\"u}rst, J.U.; Werner, C.; Beckmann, T.; Buse, K.;
  Breunig, I.
\newblock {Broadband infrared spectroscopy using optical parametric oscillation
  in a radially-poled whispering gallery resonator}.
\newblock {\em Opt. Express} {\bf 2015}, {\em 23},~24042--24047.

\bibitem[Mohageg \em{et~al.}(2005)Mohageg, Strekalov, Savchenkov, Matsko,
  Ilchenko, and Maleki]{Mohageg:2005cx}
Mohageg, M.; Strekalov, D.; Savchenkov, A.; Matsko, A.; Ilchenko, V.; Maleki,
  L.
\newblock {Calligraphic poling of Lithium Niobate}.
\newblock {\em Opt. Express} {\bf 2005}, {\em 13},~3408--3419.

\bibitem[Guarino \em{et~al.}(2007)Guarino, Poberaj, Rezzonico, Degl'Innocenti,
  and G{\"u}nter]{Guarino:2007bv}
Guarino, A.; Poberaj, G.; Rezzonico, D.; Degl'Innocenti, R.; G{\"u}nter, P.
\newblock {Electro-optically tunable microring resonators in lithium niobate}.
\newblock {\em Nature Photon.} {\bf 2007}, {\em 1},~407--410.

\bibitem[Wang \em{et~al.}(2014)Wang, Burek, Lin, Atikian, Venkataraman, Huang,
  Stark, and Lon{\v c}ar]{Wang:2014bq}
Wang, C.; Burek, M.J.; Lin, Z.; Atikian, H.A.; Venkataraman, V.; Huang, I.C.;
  Stark, P.; Lon{\v c}ar, M.
\newblock {Integrated high quality factor lithium niobate microdisk
  resonators}.
\newblock {\em Opt. Express} {\bf 2014}, {\em 22},~30924--30933.

\bibitem[Liang \em{et~al.}(2017)Liang, Lin, Luo, Jiang, Zhang, and
  Sun]{Liang:2017bz}
Liang, H.; Lin, Q.; Luo, R.; Jiang, W.C.; Zhang, X.C.; Sun, X.
\newblock {Nonlinear optical oscillation dynamics in high-Q lithium niobate
  microresonators}.
\newblock {\em Opt. Express} {\bf 2017}, {\em 25},~13504--13516.

\bibitem[Wu \em{et~al.}(2018)Wu, Zhang, Yao, Fang, Qiao, Chai, Lin, and
  Cheng]{Wu:2018bb}
Wu, R.; Zhang, J.; Yao, N.; Fang, W.; Qiao, L.; Chai, Z.; Lin, J.; Cheng, Y.
\newblock {Lithium niobate micro-disk resonators of quality factors above
  $10^7$}.
\newblock {\em Opt. Lett.} {\bf 2018}, {\em 43},~4116--4119.

\bibitem[Helmy \em{et~al.}(2010)Helmy, Abolghasem, Stewart~Aitchison, Bijlani,
  Han, Holmes, Hutchings, Younis, and Wagner]{Helmy:2010jy}
Helmy, A.S.; Abolghasem, P.; Stewart~Aitchison, J.; Bijlani, B.J.; Han, J.;
  Holmes, B.M.; Hutchings, D.C.; Younis, U.; Wagner, S.J.
\newblock {Recent advances in phase matching of second-order nonlinearities in
  monolithic semiconductor waveguides}.
\newblock {\em Laser {\&} Photon. Rev.} {\bf 2010}, {\em 5},~272--286.

\bibitem[Kuo \em{et~al.}(2014)Kuo, Bravo-Abad, and Solomon]{Kuo:2014aa}
Kuo, P.S.; Bravo-Abad, J.; Solomon, G.S.
\newblock {Second-harmonic generation using $\bar{4}$-quasi-phasematching in a
  GaAs whispering-gallery-mode microcavity.}
\newblock {\em Nat. Commun.} {\bf 2014}, {\em 5},~3109.

\bibitem[Mariani \em{et~al.}(2014)Mariani, Andronico, Lema{\^\i}tre, Favero,
  Ducci, and Leo]{Mariani:2014gw}
Mariani, S.; Andronico, A.; Lema{\^\i}tre, A.; Favero, I.; Ducci, S.; Leo, G.
\newblock {Second-harmonic generation in AlGaAs microdisks in the telecom
  range}.
\newblock {\em Opt. Lett.} {\bf 2014}, {\em 39},~3062--3064.

\bibitem[Parisi \em{et~al.}(2017)Parisi, Morais, Ricciardi, Mosca, Hansson,
  Wabnitz, Leo, and De~Rosa]{Parisi:2017fp}
Parisi, M.; Morais, N.; Ricciardi, I.; Mosca, S.; Hansson, T.; Wabnitz, S.;
  Leo, G.; De~Rosa, M.
\newblock {AlGaAs waveguide microresonators for efficient generation of
  quadratic frequency combs}.
\newblock {\em J. Opt. Soc. Am. B} {\bf 2017}, {\em 34},~1842--1847.

\bibitem[Timurdogan \em{et~al.}(2017)Timurdogan, Poulton, Byrd, and
  Watts]{Timurdogan:2017jg}
Timurdogan, E.; Poulton, C.V.; Byrd, M.J.; Watts, M.R.
\newblock {Electric field-induced second-order nonlinear optical effects in
  silicon waveguides}.
\newblock {\em Nature Photon.} {\bf 2017}, {\em 11},~200--206.

\bibitem[Herr \em{et~al.}(2014)Herr, Brasch, Jost, Wang, Kondratiev,
  Gorodetsky, and Kippenberg]{Herr:2014ip}
Herr, T.; Brasch, V.; Jost, J.D.; Wang, C.Y.; Kondratiev, N.M.; Gorodetsky,
  M.L.; Kippenberg, T.J.
\newblock {Temporal solitons in optical microresonators}.
\newblock {\em Nat. Phot.} {\bf 2014}, {\em 8},~145--152.

\bibitem[Hansson \em{et~al.}(2018)Hansson, Parra-Rivas, Bernard, Leo, Gelens,
  and Wabnitz]{Hansson:2018jt}
Hansson, T.; Parra-Rivas, P.; Bernard, M.; Leo, F.; Gelens, L.; Wabnitz, S.
\newblock {Quadratic soliton combs in doubly resonant second-harmonic
  generation}.
\newblock {\em Opt. Lett.} {\bf 2018}, {\em 43},~6033--6036.

\bibitem[Villois and Skryabin(2019)]{Villois:2019kt}
Villois, A.; Skryabin, D.V.
\newblock {Soliton and quasi-soliton frequency combs due to second harmonic
  generation in microresonators}.
\newblock {\em Opt. Express} {\bf 2019}, {\em 27},~7098--7107.

\bibitem[Erkintalo \em{et~al.}(2019)Erkintalo, Li, Parra-Rivas, and
  Leo]{Erkintalo:2019wf}
Erkintalo, M.; Li, Z.; Parra-Rivas, P.; Leo, F.
\newblock {Dynamics of Kerr-like Optical Frequency Combs Generated via
  Phase-mismatched Second-harmonic Generation}.
\newblock  2019 Conference on Lasers and Electro-Optics Europe and European
  Quantum Electronics Conference (2019). Optical Society of America,  2019, p.
  ef\_10\_2.

\bibitem[Lobanov \em{et~al.}(2020)Lobanov, Kondratiev, Shitikov, and
  Bilenko]{Lobanov:2020kt}
Lobanov, V.E.; Kondratiev, N.M.; Shitikov, A.E.; Bilenko, I.A.
\newblock {Two-color flat-top solitonic pulses in $\chi^{(2)}$ optical
  microresonators via second-harmonic generation}.
\newblock {\em Phys. Rev. A} {\bf 2020}, {\em 101},~013831.

\bibitem[Villois \em{et~al.}(2019)Villois, Kondratiev, Breunig, Puzyrev, and
  Skryabin]{Villois:2019gl}
Villois, A.; Kondratiev, N.; Breunig, I.; Puzyrev, D.N.; Skryabin, D.V.
\newblock {Frequency combs in a microring optical parametric oscillator}.
\newblock {\em Opt. Lett.} {\bf 2019}, {\em 44},~4443--4446.

\bibitem[Parra-Rivas \em{et~al.}(2019)Parra-Rivas, Gelens, and
  Leo]{ParraRivas:2019gf}
Parra-Rivas, P.; Gelens, L.; Leo, F.
\newblock {Localized structures in dispersive and doubly resonant optical
  parametric oscillators}.
\newblock {\em Phys. Rev. E} {\bf 2019}, {\em 100},~032219.

\bibitem[Menicucci \em{et~al.}(2008)Menicucci, Flammia, and
  Pfister]{Menicucci:2008ge}
Menicucci, N.C.; Flammia, S.T.; Pfister, O.
\newblock {One-Way Quantum Computing in the Optical Frequency Comb}.
\newblock {\em Phys. Rev. Lett.} {\bf 2008}, {\em 101},~130501.

\bibitem[Pfister(2020)]{Pfister:2020ck}
Pfister, O.
\newblock {Continuous-variable quantum computing in the quantum optical
  frequency comb}.
\newblock {\em J. Phys. B: Atom. Molec. Opt. Phys.} {\bf 2020}, {\em
  53},~012001.

\bibitem[Villar \em{et~al.}(2006)Villar, Martinelli, Fabre, and
  Nussenzveig]{Villar:2006ji}
Villar, A.S.; Martinelli, M.; Fabre, C.; Nussenzveig, P.
\newblock {Direct production of tripartite pump-signal-idler entanglement in
  the above-threshold optical parametric oscillator}.
\newblock {\em Phys. Rev. Lett.} {\bf 2006}, {\em 97},~140504.

\bibitem[Pfister \em{et~al.}(2004)Pfister, Feng, Jennings, Pooser, and
  Xie]{Pfister:2004fs}
Pfister, O.; Feng, S.; Jennings, G.; Pooser, R.C.; Xie, D.
\newblock {Multipartite continuous-variable entanglement from concurrent
  nonlinearities}.
\newblock {\em Phys. Rev. A} {\bf 2004}, {\em 70},~020302.

\bibitem[Guo \em{et~al.}(2005)Guo, Zou, Zhai, Zhang, and Gao]{Guo:2005cg}
Guo, J.; Zou, H.; Zhai, Z.; Zhang, J.; Gao, J.
\newblock {Generation of continuous-variable tripartite entanglement using
  cascaded nonlinearities}.
\newblock {\em Phys. Rev. A} {\bf 2005}, {\em 71},~034305.

\bibitem[Pennarun \em{et~al.}(2007)Pennarun, Bradley, and
  Olsen]{Pennarun:2007hl}
Pennarun, C.; Bradley, A.S.; Olsen, M.K.
\newblock {Tripartite entanglement and threshold properties of coupled
  intracavity down-conversion and sum-frequency generation}.
\newblock {\em Phys. Rev. A} {\bf 2007}, {\em 76},~063812.

\bibitem[He \em{et~al.}(2017)He, Sun, Hu, Zhang, Chen, and Wang]{He:2017gh}
He, G.; Sun, Y.; Hu, L.; Zhang, R.; Chen, X.; Wang, J.
\newblock {Five-partite entanglement generation between two optical frequency
  combs in a quasi-periodic $\chi^{(2)}$ nonlinear optical crystal}.
\newblock {\em Sci. Rep.} {\bf 2017}, {\em 7},~9054.

\bibitem[Raussendorf \em{et~al.}(2003)Raussendorf, Browne, and
  Briegel]{Raussendorf:2003ca}
Raussendorf, R.; Browne, D.E.; Briegel, H.J.
\newblock {Measurement-based quantum computation on cluster states}.
\newblock {\em Phys. Rev. A} {\bf 2003}, {\em 68},~022312.

\bibitem[Chembo(2016)]{Chembo:2016ki}
Chembo, Y.K.
\newblock {Quantum dynamics of Kerr optical frequency combs below and above
  threshold: Spontaneous four-wave mixing, entanglement, and squeezed states of
  light }.
\newblock {\em Phys. Rev. A} {\bf 2016}, {\em 93},~033820.

\end{thebibliography}

\end{document}